\documentclass[pdflatex,sn-nature]{sn-jnl}

\usepackage{graphicx}%
\usepackage{multirow}%
\usepackage{amsmath,amssymb,amsfonts}%
\usepackage{amsthm}%
\usepackage{mathrsfs}%
\usepackage[title]{appendix}%
\usepackage{xcolor}%
\usepackage{textcomp}%
\usepackage{manyfoot}%
\usepackage{booktabs}%
\usepackage{algorithm}%
\usepackage{algorithmicx}%
\usepackage{algpseudocode}%
\usepackage{listings}%
\usepackage{braket}
\usepackage{upgreek}
\usepackage{lineno}

\begin{document}
\title[Article Title]{Four-Photon Interference with a High-Efficiency Quantum Dot Source}

\author*[1]{\fnm{Alistair J.} \sur{Brash}}\email{a.brash@sheffield.ac.uk}

\author[1]{\fnm{Luke} \sur{Brunswick}}

\author[2]{\fnm{Mark R.} \sur{Hogg}}

\author[1]{\fnm{Catherine L.} \sur{Phillips}}

\author[2]{\fnm{Malwina A.} \sur{Marczak}}

\author[2]{\fnm{Timon L.} \sur{Baltisberger}}

\author[3]{\fnm{Sascha R.} \sur{Valentin}}

\author[3]{\fnm{Arne} \sur{Ludwig}}

\author[2]{\fnm{Richard J.} \sur{Warburton}}

\affil*[1]{\orgdiv{School of Mathematical and Physical Sciences}, \orgname{University of Sheffield}, \orgaddress{\street{Hounsfield Road}, \city{Sheffield}, \postcode{S3 7RH}, \country{United Kingdom}}}

\affil[2]{\orgdiv{Department of Physics}, \orgname{University of Basel}, \orgaddress{\street{Klingelbergstrasse 82}, \city{Basel}, \postcode{4056}, \country{Switzerland}}}

\affil[3]{\orgdiv{Faculty of Physics and Astronomy}, \orgname{Ruhr-Universit\"{a}t Bochum}, \orgaddress{\street{Universit\"{a}tsstrasse 150}, \city{Bochum}, \postcode{44801}, \country{Germany}}}

\abstract{
While two-photon Hong-Ou-Mandel interference visibility has become a standard metric for single-photon sources, many optical quantum technologies require the generation and manipulation of larger photonic states.
To date, efficiency limitations have prevented scaling quantum dot-based interference to the coalescence of more than two photons at a single beamsplitter.
We overcome this limitation by combining a state-of-the-art quantum dot source with deterministic demultiplexing, enabling the direct observation of quantum interference fringes arising from up to four photons.
We measure high mean interference contrasts of $93.0 \pm 0.1~\%$ for two photons, and $84.1 \pm 1.0~\%$ for four photons, with the complex fringe structure fully reproduced by a theoretical model. 
These results reveal the existence of  ``deep fringes'' whose minima are unaffected by distinguishable photons, rendering the maximum contrast of four-photon interference highly sensitive to multi-photon emission but robust against photon distinguishability.
We predict that these phenomena will extend to interference of larger numbers of photons, with relevance across a range of potential optical quantum technologies.
A Fisher information analysis demonstrates that interference fringes from our source can exhibit phase sensitivity beyond the standard quantum limit, illustrating potential applications in quantum metrology. 
}

\maketitle

\section*{Introduction}\label{sec:intro}

Quantum dots (QDs) integrated within photonic nanostructures are a leading platform for generating quantum light, providing a robust route to on-demand single photons with high brightness, purity, and indistinguishability \cite{Liu2018,PhysRevLett.126.233601,doi:10.1126/sciadv.abc8268,tomm_bright_2021,Ding2025}. While photon indistinguishability is typically quantified using two-photon Hong-Ou-Mandel (HOM) interference, this effect is also a cornerstone of optical quantum technologies, enabling entanglement generation and logic gates with photonic qubits \cite{PhysRevLett.95.010501}. However, the two-photon case represents the simplest instance of the much broader phenomenon of quantum interference. Moving beyond two photons, the direct relationship between interference visibility and photon indistinguishability breaks down, revealing a rich and complex landscape \cite{doi:10.1073/pnas.1206910110,PhysRevX.5.041015,PhysRevLett.118.153602,PhysRevLett.118.153603,Ferreri_2020,PhysRevLett.125.123603,Ma_2025}. This multi-photon interference can exhibit non-intuitive behaviours, such as a non-monotonic relationship with photon distinguishability \cite{doi:10.1073/pnas.1206910110} or the persistence of interference even among pairwise-distinguishable photons \cite{PhysRevLett.125.123603}. Understanding and harnessing this complexity is crucial, as it underpins transformative applications ranging from quantum information processing (QIP) and simulation \cite{PhysRevLett.123.250503,Sturges2021,Maring2024} to metrology \cite{doi:10.1126/science.1138007,Matthews2016}, imaging \cite{Defienne2024}, and secure communication \cite{PhysRevA.59.1829,Huang2021photonicquantumdata}.

Realising these applications requires efficient sources of multi-photon states, which has been a persistent challenge. Traditional sources rely on probabilistic processes such as spontaneous parametric down-conversion (SPDC), which can lead to low efficiencies. While QDs offer a near-deterministic alternative, creating multiple photons suitable for interference is not straightforward. Photon pair sources using the QD biexciton state are well-established \cite{PhysRevLett.122.113602}, but produce photons that are spectrally and temporally distinguishable.
The common alternative strategy, time-to-space demultiplexing of a QD single-photon source (SPS), has enabled experiments such as photonic QIP \cite{PhysRevLett.123.250503,Maring2024}, heralded generation of entangled 3- and 4-photon Greenberger-Horne-Zeilinger states \cite{PhysRevLett.132.130604, Pont2024}, determination of bounds on 4- and 6-photon indistinguishability through pairwise interference \cite{PhysRevX.12.031033, Maring2024} and ``indistinguishability purification'' of two photons \cite{PhysRevLett.133.033604}. However, these experiments distribute $N$ photons across $>N$ spatial modes, and are predominantly restricted to sequential, pairwise two-photon interference.
This leaves the core physics of $N$-photon interference at a single beamsplitter unexplored within the realm of QD photon sources.

In this study, we directly address this challenge by interfering up to four photons from our QD source at a single beamsplitter. By combining a state-of-the-art SPS based upon a QD in an open microcavity \cite{tomm_bright_2021} with deterministic demultiplexing and linear optics, we realise an efficient source of higher-order photon number states. 
Our architecture achieves order-of-magnitude increases in detected $N$-photon rates compared to equivalent experiments using SPDC sources \cite{doi:10.1126/science.1138007}, with the resulting signal-to-noise enhancement revealing a previously unobserved harmonic structure within the four-photon interference fringes. 
To explain these observations, we develop a comprehensive theoretical model that accurately reproduces our results, showing that this structure arises from complex interplay between multi-photon emission, photon distinguishability, and system losses. 
This analysis reveals a key finding: in stark contrast to the two-photon case, the maximum four-photon interference contrast is remarkably robust against photon distinguishability.
However, the sensitivity to multi-photon events is elevated compared to the two-photon case.
To explore the implications of these results, we perform a Fisher information (FI) analysis, revealing that for appropriate detection schemes, our source can surpass the standard quantum limit for phase sensitivity, opening potential applications in optical quantum metrology.

\section*{Results}\label{sec:results}

\subsection*{Theory of Multi-Photon Interference}

We begin by considering the case where up to 2 photons are incident at either of the inputs ($a/b$) of an ideal Mach-Zehnder interferometer (\autoref{fig:schematic}(a)). In the first instance, we consider a completely ideal system, with mutually indistinguishable Fock-state inputs and an interferometer free of any losses (zero coupling to mode $g$ in \autoref{fig:schematic}(a)). We define $P_{ij \rightarrow mn}$ as the probability that for an input state of $i$ and $j$ photons in input modes $a$ and $b$ respectively, we detect $m$ and $n$ photons in output modes $e$ and $f$ respectively. Beginning with the single-photon input case:

\begin{equation}
P_{10 \rightarrow 10} = \frac{1 - \mathrm{cos(\phi)}}{2}.
\label{eq:P10}
\end{equation}

If we now consider single photons incident at both inputs, HOM interference between the indistinguishable photons means that the intermediate state in the interferometer (with two spatial modes $c/d$) is the maximally path entangled two-photon ``NOON state'' $\frac{1}{\sqrt{2}}\{\ket{2_c 0_d} + \ket{0_c 2_d}\}$ \cite{PhysRevA.63.063407,doi:10.1080/00107510802091298}. Detection of coincidences between  single photons at both outputs then yields \cite{KIM199837}

\begin{equation}
P_{11 \rightarrow 11} = \frac{1 + \mathrm{cos(2\phi)}}{2},
\label{eq:P11}
\end{equation}
where the frequency of the cosine term is now doubled. This effect, whereby the $N$-photon wavepacket can be considered to have a de Broglie wavelength $\lambda/N$, with a corresponding $N$-fold increase in the interference frequency, is termed super-resolution \cite{PhysRevLett.74.4835, PhysRevLett.82.2868}.

If both photons are instead injected at the same input port, the intermediate state becomes $\frac{1}{\sqrt{2}}\ket{1_c 1_d} + \frac{1}{\sqrt{4}}\{\ket{2_c 0_d} + \ket{0_c 2_d}\}$ as there is no longer HOM interference at the first beamsplitter. This $\ket{1_c 1_d}$ component undergoes HOM interference at the second beamsplitter, meaning that it does not contribute to the coincidence signal, which is now \cite{KIM199837}

\begin{equation}
P_{20 \rightarrow 11} = \frac{1}{2}\left(\frac{1 - \mathrm{cos(2\phi)}}{2}\right).
\label{eq:P20}
\end{equation}
This is identical to the $\ket{1_a 1_b}$ input case aside from a $\pi/2$ phase shift and an additional prefactor of the ``intrinsic efficiency'' $\eta_i = \frac{1}{2}$ due to the undetected $\ket{1_c 1_d}$ component.

Finally, we now consider the four-photon input case $\ket{2_a 2_b}$. Here, the intermediate state is $\frac{1}{4} \left( \sqrt{6} \left\{ \ket{4_c 0_d}+\ket{0_c 4_d} \right\} + 2\ket{2_c 2_d} \right)$. This corresponds to the four-photon form of an $N$-photon Holland-Burnett state, generated when two $N/2$ photon Fock states coincide at a beamsplitter \cite{PhysRevLett.71.1355}. The dominant component is a 4 photon NOON state, which carries a $\frac{3}{4}$ probability weight. This component can be isolated (using either photon number resolving or multiplexed single-photon detectors) by post-selecting for odd photon numbers (e.g. $\ket{3_e 1_f}$) at the output. This is effective since the remaining $\ket{2_c 2_d}$ term produces exclusively even-numbered outputs due to HOM interference \cite{PhysRevLett.83.959,doi:10.1126/science.1138007}. Using this approach, the coincidence signal is then

\begin{equation}
P_{22 \rightarrow 31} = \frac{3}{8}\left(\frac{1 - \mathrm{cos(4\phi)}}{2}\right),
\label{eq:P22}
\end{equation}
which exhibits 4-fold super-resolution \cite{walther2004broglie,mitchell2004super} with $\eta_i = \frac{3}{8}$. This can potentially be increased to $\eta_i = \frac{3}{4}$ through simultaneous detection of both $\ket{3_e1_f}$ and $\ket{1_e3_f}$ \cite{Okamoto_2008}. We focus our investigation on these output states as other possibilities ($\ket{4_e0_f}/\ket{0_e4_f}$ and $\ket{2_e2_f}$) do not produce fringes exhibiting consistent 4-fold super-resolution (further details in supplementary material).

To go beyond the ideal case described by these analytical expressions, we construct a versatile model based on the input-output formalism (full details in supplementary material).
This model incorporates potential experimental imperfections including multi-photon emission, photon loss, and the degree of photon indistinguishability, quantified by the parameter $\mathcal{I}$.
For ideal, indistinguishable ($\mathcal{I}=1$) Fock state inputs (dark lines in \autoref{fig:schematic}(b-d)), the model directly reproduces the analytical expressions of Eqs. \ref{eq:P10} - \ref{eq:P22}.
As a first step, we investigate the role of quantum interference by modelling fringes for fully distinguishable ($\mathcal{I}=0$) photons (pale lines in \autoref{fig:schematic}(b-d)).
While the single-photon fringes remain unaffected (being governed purely by single-particle interference), the two-photon fringes degrade: the $\ket{1_a 1_b}$ input exhibits shallower minima, whilst the $\ket{2_a 0_b}$ input shows reduced maxima. 
The most striking change occurs for the $\ket{2_a 2_b}$ input: here, the fringe amplitude is strongly suppressed and every second minimum is lost, washing out the clear $4\phi$ oscillations that are present at $\mathcal{I}=1$. 
This stark change in fringe structure underscores the rich complexity of four-photon interference, motivating our experimental investigations.

\subsection*{Preparation of Photon Number States}

\begin{figure}[ht]
\centering
\includegraphics[width=1\textwidth]{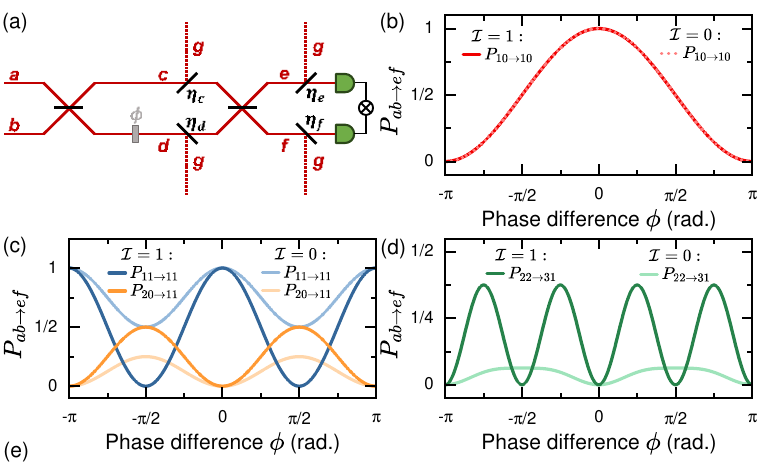}
\includegraphics[width=1\textwidth]{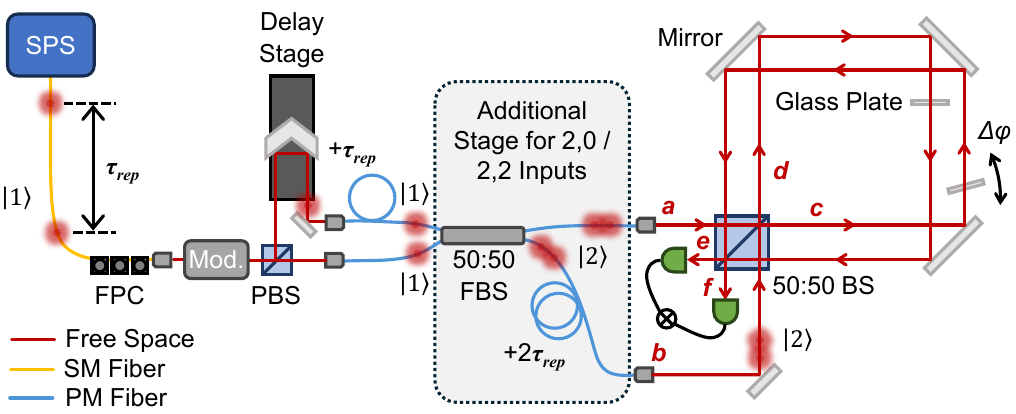}
\caption{(a) Schematic of the model of multi-photon interference: $a/b$ - input modes, $c/d$ - interferometer modes, $e/f$ - output modes, $g$ - photon loss mode , $\eta$ - transmission efficiency of the mode specified by the subscript, $\phi$ - phase shift. (b-d) Interference fringes produced by the input-output model for Fock state inputs up to 4 photons -- dark / pale lines correspond respectively to indistinguishable / distinguishable photon inputs ($\mathcal{I} = 1 / 0$). (e) Schematic of the experimental configuration. Single photons from the source are demultiplexed into two spatial modes $a/b$, with the first photon in $a$ delayed by the laser pulse separation $\tau_{\rm rep}$. An optional additional two-photon interference step (grey shaded box) combined with a further delay of $2\tau_{\rm rep}$ produces 2 photon inputs. The photons then pass through a displaced Sagnac interferometer (functionally equivalent to the schematic in (a)), where tilting a glass plate induces a phase shift $\phi$ between the two counter-propagating modes $c/d$. Photons at the output ports $e/f$ are detected by an array of up to 4 detectors, with correlations evaluated by time-tagging electronics. Solid lines: red - free space beam, yellow - single mode (SM) fibre, blue - polarization maintaining (PM) fibre. Acronyms: SPS - single-photon source, FPC - fibre polarization controller, Mod. - phase modulator, PBS - polarizing beamsplitter, FBS - fibre beamsplitter.}\label{fig:schematic}
\end{figure}

We now turn our attention to realising this multi-photon interference experimentally. \autoref{fig:schematic}(e) illustrates the principle of our measurements. Single photons are generated by the QD-microcavity source (further details in Ref. \cite{tomm_bright_2021}), operating in the ``blue collection'' configuration described in Ref. \cite{Javadi_2023}. The source is excited with a laser pulse area $\Theta = \pi$, producing a charged exciton ($X^+$) population of $0.98^{+0.01}_{-0.05}$ in the QD \cite{Javadi_2023}. We measure a source efficiency of $53 \pm 4~\%$, corresponding to an output in-fibre single photon flux of $\sim 40 ~\mathrm{MHz}$ when excited by a mode-locked laser with a repetition rate of $1/\tau_{\rm rep} = 76.3 ~\mathrm{MHz}$. Under these conditions, the measured single-photon purity is $\mathrm{g^{(2)}(0)} = 0.018 \pm 0.001$. Crucially for multi-photon interference experiments, our source maintains constant HOM interference visibility ($V_{\rm HOM}$) for photons separated by up to $1.5~\upmu \mathrm{s}$ ($> 100 \, \tau_{\rm rep}$) \cite{tomm_bright_2021}, ensuring that all photons exhibit high mutual indistinguishability.

A resonant polarization modulator demultiplexes subsequent photons into two spatial modes (see Methods). The photons are then coupled to the slow axes of polarization-maintaining (PM) fibres with an extinction ratio of $20~\mathrm{dB}$. A $\ket{1_a1_b}$ state is prepared by delaying one path by $\tau_{\rm rep}$, achieved primarily through addition of extra fibre length and tuned by a delay stage (see Methods). To increase the number of photons per spatial mode, the addition of a PM fibre beamsplitter followed by a further delay of $2\tau_{\rm rep}$ in one path (grey shaded region in \autoref{fig:schematic}(e)) can prepare a $\ket{2_a 2_b}$ state through HOM interference, with a success probability of $\frac{1}{4}$ that both photon pairs arrive at the interferometer in opposite spatial but identical temporal modes. The transmission of the state preparation setup, defined as the total excess loss between the SPS output fibre and interferometer input modes $a/b$, is approximately $54\%$ in the $\ket{11}$ configuration and $47\%$ in the $\ket{22}$ configuration (excluding the aforementioned $\frac{1}{4}$ success probability for $\ket{22}$ preparation). The losses are dominated by the fibre coupling efficiency (typically around $70\%$) after the free-space modulator.

\subsection*{Interferometry of Photon Number States}

\begin{figure}[ht]
\centering
\includegraphics{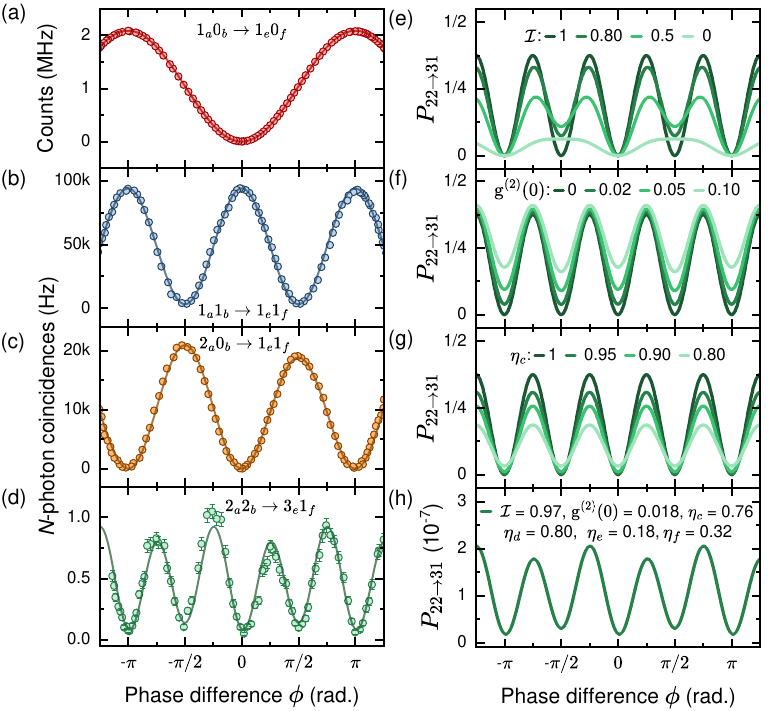}
\caption{Experimental (circles) and theoretical (solid lines) interference fringes for various $N$-photon input states as a function of the phase ($\phi$) between modes $c/d$: (a) Count rate for a single photon input. (b) Two-photon coincidence rate for single photon inputs to both ports. (c) Two-photon coincidence rate for two-photon input to a single port. (d) Four-photon coincidence rate for two-photon inputs to both ports. Parameters for the theoretical curves in (a-d) are found by fitting the experimental data (full details in supplementary material). (e-g) Theoretical four-photon interference fringes for ideal parameters with varying: (e) photon indistinguishability ($\mathcal{I}$), (f) single-photon purity $\mathrm{g^{(2)}(0)}$ and (g) interferometer mode $c$ transmission $\eta_c$ (with $\eta_d=1$). (h) Theoretical four-photon interference fringes using parameters extracted from fit to experiment in (d).}\label{fig:NPhotonFringes}
\end{figure}

To perform phase-resolved photon correlation measurements, the outputs of the state preparation stage are connected to the two input ports ($a/b$) of a displaced Sagnac interferometer \cite{10.1063/1.4891702} (\autoref{fig:schematic}(e)), with the slow axis of both input fibres aligned to s-polarisation. While functionally equivalent to the well-known Mach-Zehnder interferometer (\autoref{fig:schematic}(a)), this configuration harnesses two counter-propagating modes ($c/d$) sharing common optical surfaces to achieve very high passive phase stability \cite{10.1063/1.4891702}. The beamsplitter (BS) has a near-ideal splitting ratio of $0.505:0.495$ at the QD source wavelength of $923~\mathrm{nm}$. A relative phase shift ($\phi$) between the two modes is achieved by tilting a glass plate in one arm relative to the optical axis (see Methods). Photons at the two output ports ($e/f$) are coupled to single mode fibre and detected by superconducting nanowire single-photon detectors (SNSPDs), with time-tagging electronics allowing evaluation of correlations between detected photons (further details in Methods).

To characterize the behaviour of multi-photon interference from our source, we perform a series of single-, two- and four-photon interference experiments.
We begin our investigation with the single-photon case in \autoref{fig:NPhotonFringes}(a), blocking input $b$ to leave single photons incident only on input $a$. As expected from Equation \ref{eq:P10}, the SNSPD counts at output $e$ oscillate sinusoidally as a function of the phase plate angle $\theta_p$. As the phase plate has a finite surface flatness ($\lambda / 10$), the relationship between its angle $\theta_p$ and the interferometer phase $\phi$ is not perfectly linear. To account for this, we calibrate the experimental relationship between $\theta_p$ and $\phi$ using the single-photon fringe data as a phase reference (full details in supplementary material).

The contrast of the experimental oscillations can be evaluated according to:

\begin{equation}
C = \frac{I_{\rm max}-I_{\rm min}}{I_{\rm max}+I_{\rm min}},
\end{equation}
where $I_{\rm max}$ and $I_{\rm min}$ are the count rates detected at the maxima and minima of the fringe pattern respectively. To minimise uncertainty, we consider an average contrast $\bar{C}$ over multiple adjacent maxima and minima within our $2 \pi$ phase scan, calculating the uncertainty through Gaussian error quadrature. Applying this to the single-photon input in \autoref{fig:NPhotonFringes}(a), the maximum ($2.08~\mathrm{MHz}$) and minimum ($530~\mathrm{Hz}$) observed count rates yield a single-photon interference contrast of $\bar{C}_{10} = 0.999 \pm 0.001$, with similar results observed for all four single-photon permutations of inputs ($a/b$) and outputs ($e/f$).
The solid line is a fit of our input-output formalism model (full parameters for all fits included in supplementary material) which shows excellent agreement with the experimental data.

We now consider the two-photon case, inputting one photon simultaneously to both $a$ and $b$ and detecting coincident SNSPD clicks at $e$ and $f$. For all photon correlation measurements, a $2~\mathrm{ns}$ coincidence window is specified. This greatly exceeds the QD source's radiative lifetime of $T_1 = 59 \pm 3~\mathrm{ps}$ but is also significantly less than the laser pulse separation $\tau_{\rm rep} = 13.2~\mathrm{ns}$, ensuring that we capture all coincidences from a single cycle of the experiment without any spurious events.
The results of this experiment are shown in \autoref{fig:NPhotonFringes}(b), illustrating a doubling of the oscillation frequency compared to the single-photon data in \autoref{fig:NPhotonFringes}(a), in line with \autoref{eq:P11}. By comparing the maximum and minimum coincidence rates, we extract a contrast $\bar{C}_{11} = 0.930 \pm 0.001$ and a maximum detected two-photon coincidence rate of $94.0 \pm 0.1~\mathrm{kHz}$, exceeding equivalent previous measurements with SPDC \cite{doi:10.1126/science.1138007} or QD \cite{PhysRevLett.118.257402} sources by 1 and 4 orders of magnitude, respectively. We note that our value of $\bar{C}_{11}$ is comparable to the raw HOM interference visibilities ($V_{\rm HOM}$) obtained for our QD source in previous work \cite{tomm_bright_2021}, implying that this is the limiting parameter for $\bar{C}_{11}$. Our model again produces excellent agreement with the experimental data (solid line), with the critical fitted parameter of photon indistinguishability $\mathcal{I} = 0.974 \pm 0.07$ (defined as the mean photon wavepacket overlap of the single-photon emission \cite{PhysRevLett.126.063602}) consistent with previous values of $\mathcal{I}$ measured for our QD source by correcting $V_{\rm HOM}$ for finite $\mathrm{g^{(2)}(0)}$ \cite{tomm_bright_2021}.

Exploiting the versatility of our state preparation apparatus, we now consider a different two-photon input state. By adding the additional 50:50 fibre beamsplitter after the modulator (grey shaded region in \autoref{fig:schematic}(e)), HOM interference of indistinguishable photons leads to the preparation of 2 photon states in the input modes $a/b$. By blocking input $b$, we create a $\ket{2_a 0_b}$ input state. 
\autoref{fig:NPhotonFringes}(c) shows the results of this experiment. As predicted, the oscillations exhibit the same period as \autoref{fig:NPhotonFringes}(b), with the peak coincidence rate of $21.0 \pm 0.1~\mathrm{kHz}$ approximately one quarter of that observed in \autoref{fig:NPhotonFringes}(b). The reduced coincidence rate is due to the intrinsic efficiency $\eta_i = \frac{1}{2}$ described in \autoref{eq:P20}, the loss of half the input photons at the blocked input mode $b$, and the slightly reduced throughput of our state preparation apparatus in the two-photon configuration. A contrast of $\bar{C}_{20} = 0.988 \pm 0.002$ is evaluated, significantly higher than the $\ket{1_a 1_b}$ input.

Interestingly, the $\ket{2_a 0_b}$ fringes exhibit an asymmetry between neighbouring maxima not seen for single-photon inputs (\autoref{fig:NPhotonFringes}(a,b)). This corresponds to a modulation at half the frequency of the $2\phi$ oscillation predicted by \autoref{eq:P20}. The theoretical fit confirms that this arises from coincidences between one photon from the target $\ket{2_a 0_b}$ state and an extra distinguishable photon from a multi-photon emission event. Physically, this effect requires both non-zero $\mathrm{g^{(2)}(0)}$ and different channel efficiencies (e.g., $\eta_c \neq \eta_d$ or $\eta_e \neq \eta_f$), with the magnitude predominantly determined by $\mathrm{g^{(2)}(0)}$. In contrast, the $\ket{1_a 1_b}$ input remains symmetric: this stems from the balanced input state and equal multi-photon probabilities in modes $a$ and $b$.

We now unblock path $b$ to prepare the full $\ket{2_a2_b}$ input state.
To implement the $\ket{3_e 1_f}$ detection scheme, we split output $e$ across three individual SNSPDs using fibre beamsplitters. This achieves a quasi-photon number resolving configuration, with the overhead of a success probability $\frac{1}{9}$ that all three photons take the correct paths to produce a three photon coincidence. \autoref{fig:NPhotonFringes}(d) illustrates the results of this measurement. In accordance with our expectation for a four-photon state, the frequency of the oscillations is doubled compared with the two-photon measurements in \autoref{fig:NPhotonFringes} (b,c) and quadrupled compared with the single-photon measurement of \autoref{fig:NPhotonFringes}(a). The peak four-photon coincidence rate is $1.05 \pm 0.06 ~\mathrm{Hz}$, approximately an order of magnitude larger than comparable previous experiments with a SPDC source \cite{doi:10.1126/science.1138007}. The mean contrast $\bar{C}_{22} = 0.841 \pm 0.010$, extracted from the 8 permutations of adjacent maxima/minima in \autoref{fig:NPhotonFringes}(d), is also comparable (within uncertainties) to these previous SPDC experiments \cite{doi:10.1126/science.1138007}.

Our observed mean four-photon contrast $\bar{C}_{22} = 0.841 \pm 0.010$ approximates the square of the two-photon contrast $\bar{C}_{11} = 0.930 \pm 0.001$, as one might intuitively expect. 
However, the fringes exhibit significant structure that is not captured by a simple average contrast. In addition to the primary $4 \phi$ oscillations predicted in \autoref{eq:P22}, the amplitudes of both the maxima and minima oscillate at half this frequency, with the ``deepest'' adjacent maxima / minima pairing having a contrast of $C^{\rm max}_{22} = 0.909 \pm 0.023$, whilst the ``shallowest'' pairing yields a value of $C^{\rm min}_{22} = 0.778 \pm 0.038$. Our input-output model (solid line in \autoref{fig:NPhotonFringes}(d)) fully reproduces the complex features of the experimental fringe pattern with realistic parameters (listed in supplementary material).

To explore the origin of this additional harmonic structure, \autoref{fig:NPhotonFringes}(e-g) plots modelling of ideal four-photon interference fringes ($P_{22 \rightarrow 31}$) as key parameters are varied. As detailed previously (\autoref{fig:schematic}(d)), reducing photon indistinguishability to $\mathcal{I}=0$ washes out the $4 \phi$ oscillations. \autoref{fig:NPhotonFringes}(e) reveals that as $\mathcal{I}$ decreases, this occurs through a reduction in maxima height combined with the raising of every second minimum. Crucially, however, the ``deep'' minima at integer $\pi$ phases ($\phi = k \pi, k \in \mathbb{Z}$) remain unaffected, corresponding to maxima in the $\ket{2_e2_f}$ output probability (see supplementary material). To gain insight, we look to the two-photon experiments at these same phase values: we observe maxima in $P_{11 \rightarrow 11}$ and minima in $P_{20 \rightarrow 11}$ (\autoref{fig:NPhotonFringes}(b-c)). This indicates that at $\phi = k \pi$, the interferometer reproduces the input photon statistics. Consequently, $P_{22 \rightarrow 31}$ is suppressed at these points due to the specific parity of the input (even) and detected (odd) photon numbers, regardless of photon indistinguishability.

Examining other parameters, \autoref{fig:NPhotonFringes}(f) reveals that imperfect single-photon purity ($\mathrm{g^{(2)}(0)} > 0$) primarily raises the minima. By contrast, introducing asymmetric loss in the interferometer ($\eta_c < \eta_d = 1$, see \autoref{fig:NPhotonFringes}(g)) uniformly degrades contrast, whereas non-unity detector efficiencies ($\eta_e , \eta_f < 1$) have no impact on contrast under otherwise ideal conditions and are therefore not plotted.

Critically, none of these individual mechanisms can alone reproduce the full experimental structure shown in \autoref{fig:NPhotonFringes}(d); specifically, none produce the observed oscillation in maxima heights. Qualitative agreement is only achieved by modelling the interplay of realistic parameters (\autoref{fig:NPhotonFringes}(h)). Here, the degradation of the minima at $\phi = \frac{k\pi}{2}$ is driven by reduced indistinguishability ($\mathcal{I} < 1$). The remaining complex features, such as the oscillating peak heights, arise when at least one photon from the target $\ket{2_a2_b}$ input is lost and the coincidence detection is completed by distinguishable photons from multi-photon emission events. Specifically, in the presence of finite $\mathrm{g^{(2)}(0)} > 0$, symmetric efficiency reductions ($\eta_c =\eta_d$ or $\eta_e =\eta_f$) lower the height of alternating fringes, while increasing interferometer asymmetry ($|\eta_c - \eta_d|$) uniformly suppresses all peaks. Finally, increasing detector asymmetry ($|\eta_e - \eta_f|$) specifically degrades the $\phi=k\pi$ minima, which are notably distinct from those affected by $\mathcal{I}$.

Experimentally, the high efficiency of our source is key to reducing the individual uncertainties to the point where this additional fringe structure can be resolved. We expect similar features would also be present in the parameter regime of previous SPDC experiments \cite{doi:10.1126/science.1138007}; however, as the uncertainties were comparable to the four-photon signal amplitude at the fringe minima, these fine features were likely obscured, leading to the use of a simple sinusoidal fit to extract the contrast. The variables in our model exhibit significant mutual dependencies, which would lead to large uncertainties in a full free-parameter fit to the four-photon fringes. However, by fixing independently measured parameters such as $\mathrm{g^{(2)}(0)}$ and incorporating values from single- and two-photon fits, we obtain a single consistent parameter set that describes all experiments with low uncertainties (see supplementary material for full details).

\subsection*{Influence of Source Properties}

\begin{figure}[ht]
\centering
\includegraphics[width=1\textwidth]{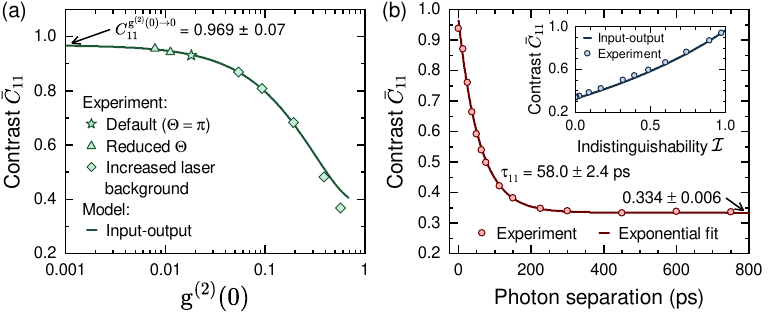}
\caption{Variation of the mean two-photon interference contrast $\bar{C}_{11}$ with source parameters. (a) $\bar{C}_{11}$ as a function of $\mathrm{g^{(2)}(0)}$. Solid symbols denote experimental data: star - default configuration used in Fig. \ref{fig:NPhotonFringes}, triangles - reduced laser pulse area $\Theta$, diamonds - increased laser background leakage. Solid line - input/output model. (b) $\bar{C}_{11}$ as a function of photon separation. Solid circles denote experimental data, with the solid line showing an exponential fit. Inset: data from main figure (blue circles) plotted against the input-output model prediction (blue line) by converting photon separation to indistinguishability (see supplementary material for details).}\label{fig:G2andTdep}
\end{figure}

We now turn our attention to experimental investigations of the source parameters that limit the contrasts measured in the previous section. To do so, we perform additional measurements of $\bar{C}_{11}$ as a function of $\mathcal{I}$ and $\mathrm{g^{(2)}(0)}$, with the high two-photon coincidence rates enabling low uncertainties for short measurement durations. We begin by investigating the influence of multi-photon emission probability ($\mathrm{g^{(2)}(0)}$) in \autoref{fig:G2andTdep}(a). The default condition used for the experiments in \autoref{fig:NPhotonFringes}(a-d) is indicated by the star-shaped marker in \autoref{fig:G2andTdep}(a), corresponding to a laser pulse area $\Theta = \pi$ and a signal-to-background ratio (SBR) of around $380:1$ between QD fluorescence and scattered laser photons from the resonant excitation.

From here, we are able to reduce $\mathrm{g^{(2)}(0)}$ by reducing the excitation pulse area first to $\Theta = \frac{3\pi}{4}$ and then $\frac{\pi}{2}$ (triangular markers in \autoref{fig:G2andTdep}(a)). The reduction of $\Theta$ decreases the probability that the QD can be re-excited by the same laser pulse after emission of a single photon \cite{Fischer2017}, reducing $\mathrm{g^{(2)}(0)}$ from $0.018 \pm 0.001$ to $0.008 \pm 0.001$, though this comes at the expense of halving the source brightness due to the reduced excited state population. With this reduced $\mathrm{g^{(2)}(0)}$, $\bar{C}_{11}$ increases from $0.930 \pm 0.001$ to $0.955 \pm 0.001$, demonstrating that the presence of additional photons from a finite $\mathrm{g^{(2)}(0)}$ degrades the observed two-photon interference fringes. To access larger values of $\mathrm{g^{(2)}(0)}$, we instead artificially reduce the SBR of the source, allowing additional laser photons to leak into the collection fibre by reducing the polarisation rejection ratio (diamond symbols in \autoref{fig:G2andTdep}(a)). The largest value $\mathrm{g^{(2)}(0)} = 0.569 \pm 0.001$ is reached at a SBR of approximately 2:1, with a corresponding large drop in $\bar{C}_{11}$ to $0.367 \pm 0.016$ again indicating an inverse relationship between $\bar{C}_{11}$ and $\mathrm{g^{(2)}(0)}$.

Our input-output model (green line in \autoref{fig:G2andTdep}(a)) provides good agreement with the experimental data using the parameters extracted from the fit to \autoref{fig:NPhotonFringes}(b).
Extrapolating our model towards $\mathrm{g^{(2)}(0)} = 0$ yields a maximum contrast for perfect single-photon purity of  $C_{11}^{\mathrm{g^{(2)}(0)}  \rightarrow 0} = 0.969 \pm 0.007$, indicating that the value of $\bar{C}_{11} = 0.930\pm 0.001$ measured in \autoref{fig:NPhotonFringes}(b) is limited by approximately equal contributions from finite $\mathrm{g^{(2)}(0)}$ and residual QD decoherence sources, likely phonon scattering \cite{IlesSmith2017,PhysRevLett.123.167403} due to the short timescale probed here ($\tau_{\rm rep} = 13.2~\mathrm{ns}$).
At large $\mathrm{g^{(2)}(0)} > 0.2$, the model diverges from the experimental points due the breakdown of the intrinsic assumption that $\mathrm{g^{(2)}(0)} \ll 1$ for a single-photon source.

We note that our model explicitly assumes that multi-photon events consist of two photons: one indistinguishable from the single-photon component and one fully distinguishable. This aligns with the expression $\mathcal{I} = \left(1 + 2\mathrm{g^{(2)}(0)}\right)V_{\rm HOM}$, often applied to estimate indistinguishability from HOM interference measurements of QDs \cite{Santori2002,tomm_bright_2021}.
However, recent work \cite{PhysRevLett.126.063602,gonzalezruiz2024} indicates that, rather than a fixed value of 2, the prefactor ($F$) to the $\mathrm{g^{(2)}(0)}$ term lies in the range $1 \leq F \leq 3$ depending upon the degree of distinguishability between various components \cite{gonzalezruiz2024}.
The strong agreement between experiments and our model in \autoref{fig:G2andTdep}(a) suggests that any deviation from $F = 2$ is small for our source parameters.

Another potential source of photon distinguishability is the temporal overlap of the two photons as they meet at the interferometer BS. Adjusting the delay of one path after the modulator controllably introduces temporal distinguishability between the two photons, with the results shown in \autoref{fig:G2andTdep}(b). The contrast falls exponentially as a function of photon separation.
The extracted time constant $\tau_{\rm 11} = 58.0 \pm 2.4 ~\mathrm{ps}$ (solid red line in \autoref{fig:G2andTdep}(b)) corresponds within uncertainty to the QD radiative lifetime ($T_1 = 59 \pm 3 ~\mathrm{ps}$) measured independently using time-resolved resonance fluorescence. Such agreement confirms that the fringe contrast follows the temporal dependence expected for HOM interference visibility where coherence is governed primarily by radiative relaxation \cite{Bylander2003}. This is consistent with the high maximum contrast (and underlying value of $\mathcal{I}$) in \autoref{fig:G2andTdep}(a), indicating that residual pure dephasing contributions are small.

Notably, even for photon separations $\gg T_1$, $\bar{C}_{11}$ plateaus at $0.334 \pm 0.006$. This agrees with our input-output model, which predicts a minimum contrast of $1/3$ for completely distinguishable photons, demonstrating that super-resolution fringes persist even in the absence of two-photon interference. We attribute this to the summation of independent single-photon probabilities: since the photons are uncorrelated, the total coincidence probability is the sum of the probabilities that both photons follow ``straight'' ($a \rightarrow e$, $b \rightarrow f$) or ``crossed'' ($a \rightarrow f$, $b \rightarrow e$) paths through the interferometer. For the straight paths, this is simply the square of \autoref{eq:P10}, whilst the crossed paths add a $\pi$ phase shift to \autoref{eq:P10} before squaring. The sum is then $P_{11} = \sin^4{\frac{\phi}{2}} + \cos^4{\frac{\phi}{2}} = \frac{1}{4}\left(3 + \cos{2\phi} \right)$, a $2\phi$-periodic fringe with a contrast of 1/3 (pale blue line in \autoref{fig:schematic}(c)). Similar fringes with two-fold super-resolution have been observed in SPDC experiments with temporally distinguishable photons \cite{doi:10.1126/science.1138007,Kim2016}, though with differing values of contrast due to the specific source statistics and experimental configurations.

To explicitly relate the results of \autoref{fig:G2andTdep}(b) to photon indistinguishability, we map photon separation to temporal indistinguishability by calculating the wavefunction overlap (see supplementary material). The resulting data is displayed in the inset to \autoref{fig:G2andTdep}(b). The excellent agreement between the transformed experimental data (blue circles) and our model (blue line) confirms that the degradation of interference contrast is driven specifically by the increasing temporal distinguishability of the photons.

\subsection*{Contrast \& Phase Sensitivity of Multi-Photon Interference}
\begin{figure}
    \centering
    \includegraphics[]{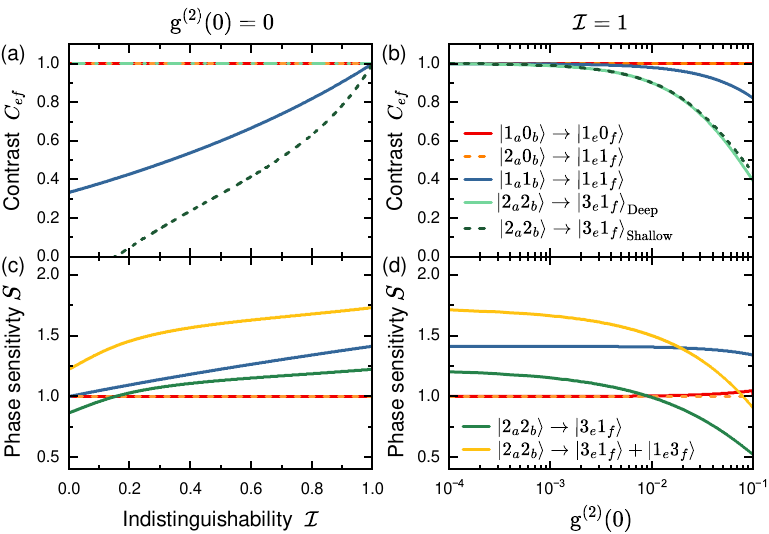}
    \caption{Calculated variation of (a,b) interference contrast $C_{ef}$ and (c,d) maximum phase sensitivity $S$ from our model as a function of source photon indistinguishability $\mathcal{I}$ (a,c) and multi-photon emission probability $\mathrm{g^{(2)}(0)}$ (b,d). Where multiple lines overlap, dashes are used. As four-photon fringes $\ket{2_a 2_b} \rightarrow \ket{3_e 1_f}$ in general exhibit alternating deep/shallow fringes (see \autoref{fig:NPhotonFringes}), their respective contrasts are plotted separately as bright/dark green lines respectively in (a,b). Aside from the variable under investigation, model parameters are set to ideal values: $\eta_{c,d,e,f} = 1$, (a,c) $\mathrm{g^{(2)}(0) = 0}$, (b,d) $\mathcal{I} = 1$.}
    \label{fig:paramsweeps}
\end{figure}

While two-photon interference contrast / visibility ($C_{11}/V_{\rm HOM}$) directly maps to indistinguishability ($\mathcal{I}$ -- see \autoref{fig:G2andTdep}(b)), our experimental four-photon interference fringes (\autoref{fig:NPhotonFringes}(d)) and modelling (\autoref{fig:NPhotonFringes}(f)) reveal a more nuanced relationship, where reduced $\mathcal{I}$ selectively impacts only every second interference minima. This suggests that the relationship between $N > 2$ photon interference and $\mathcal{I}$ presents an interesting avenue for further study.
To do this, we apply our input-output model to calculate the influence of $\mathcal{I}$ on the multi-photon interference contrast $C_{ef}$ for otherwise ideal parameters (\autoref{fig:paramsweeps}(a)).
We account for the aforementioned alternating minima depth in four-photon fringes with non-unity $\mathcal{I}$ (\autoref{fig:NPhotonFringes}(d,f)) through the nomenclature of ``deep / shallow'' fringes (pale / dark green lines respectively in \autoref{fig:paramsweeps}(a,b)).
We observe that single mode inputs ($\ket{1_a 0_b}$ / $\ket{2_a 0_b}$ -- red / orange lines) are unaffected by $\mathcal{I}$, whilst the two-photon interference contrast ($\ket{1_a 1_b}$ -- blue line) reduces considerably with $\mathcal{I}$, reaching a value of $C_{11}=1/3$ at $\mathcal{I}=0$ that matches the right hand side of \autoref{fig:G2andTdep}(b).
Moving to the four-photon case, the ``shallow'' fringes (dashed dark green line) decay significantly faster than the two-photon case, however the ``deep'' fringes (light green line) retain high contrast, illustrating the non-trivial relationship between contrast and $\mathcal{I}$ for the four-photon case.

Repeating this analysis for the relationship with $\mathrm{g^{(2)}(0)}$ (\autoref{fig:paramsweeps}(b)) shows that four-photon interference (green lines) is markedly more sensitive to multi-photon events than two-photon measurements (blue line).
Indeed, we observe that in the parameter regimes typical of resonantly driven QD SPSs, the detrimental influence of finite $\mathrm{g^{(2)}(0)}$ is much stronger than that of non-unity $\mathcal{I}$.

We now turn our attention to analysing the phase sensitivity of these multi-photon interference fringes with a view to metrology applications.
For experiments using a classical field with mean photon number $N$, the precision of the phase estimation cannot exceed the standard quantum limit (SQL) of $\Delta\phi_{\rm SQL} = 1/\sqrt{N}$. However, for $N$ entangled particles, this limit instead becomes the Heisenberg limit of $\Delta\phi_{\rm HL} = 1/N$.
Using the framework of Fisher information (FI) \cite{doi:10.1142/S0219749909004839}, we can then define a phase sensitivity $S = 1$ for the SQL, with the Heisenberg limit being $S=\sqrt{N}$ and any $S>1$ exhibiting ``super-sensitivity'' beyond the classical limit \cite{Okamoto_2008}.
We note that the definition of super-sensitivity is conditional upon the accounting of the resources used \cite{PhysRevLett.98.223601}. As we are focused on a novel source implementation, we exclude losses associated with the interferometer and detectors from our sensitivity analysis, as is typical in prior works \cite{Okamoto_2008,doi:10.1126/science.1138007,PhysRevLett.118.257402,Ma_2025}.

Applying the FI analysis (full details in supplementary material) allows us to extract the maximum phase sensitivity ($S$) of the two-photon fringes in \autoref{fig:NPhotonFringes}(b) as $S_{11 \rightarrow 11}=1.39$, approaching the Heisenberg limit of $\sqrt{2}$. This is a significant advance in sensitivity upon previous QD-based demonstrations ($S = 1.23$ \cite{PhysRevLett.118.257402}, $S = 1.13$ \cite{photonics11060512}) \footnote{Values of $S$ for previous works are calculated from reported $N$-photon fringe contrasts according to the approach described in Ref. \cite{Okamoto_2008}}, noting that the detected coincidence rate is also increased by 4 orders of magnitude \cite{PhysRevLett.118.257402}. Meanwhile, comparable previous SPDC experiments achieved a similar $S = 1.36$, but with a $\sim 15 \times$ lower coincidence rate. Repeating for the four-photon fringes in \autoref{fig:NPhotonFringes}(d) yields $S_{22 \rightarrow 31}=0.91$, lying below both the two-photon case and the SQL due to the low intrinsic efficiency $\eta_i=\frac{3}{8}$ of the $\ket{3_e1_f}$ detection scheme. We note that applying the same analysis to an equivalent SPDC experiment \cite{doi:10.1126/science.1138007} also results in $S_{22 \rightarrow 31} < 1$ for similar reasons \cite{Okamoto_2008}. To overcome this, simultaneous detection of $\ket{3_e1_f}/\ket{1_e3_f}$ would increase $\eta_i$ to $\frac{3}{4}$ \cite{Okamoto_2008}, yielding $S_{22 \rightarrow 31+13}=1.41$ for our source parameters, compared to $S_{22 \rightarrow 31+13}=1.30$ in an equivalent SPDC experiment \cite{Okamoto_2008}. Whilst experimentally we were limited by the availability of only 4 SNSPDs, this illustrates that our present source parameters are sufficient to beat both the SQL and two-photon scheme with an appropriate four-photon detection scheme. Whilst ultimately the four-photon Heisenberg limit is $S = 2$, the intrinsic efficiency of a Holland-Burnett state cannot exceed $\eta_i = \frac{3}{4}$, resulting in an upper-bound of $S = \sqrt{3}$.

To explore the limits to $S$ in our experiments, we apply our model to plot the maximum $S$ value as a function of $\mathcal{I}$ for otherwise ideal parameters (\autoref{fig:paramsweeps}(c)), revealing strikingly different scaling behaviours for different input states. For the two-photon case (blue line), reducing $\mathcal{I}$ degrades $S$ in a near-linear manner. In contrast, the four-photon sensitivity (green and yellow lines for $\ket{3_e1_f}$ and $\ket{3_e1_f}/\ket{1_e3_f}$ detection, respectively) initially exhibits a significantly shallower gradient. A turning point occurs as $\mathcal{I}$ drops below 0.3, after which the gradient becomes significantly steeper than the two-photon case. We attribute this initial resilience to the ``deep fringes'' identified in \autoref{fig:NPhotonFringes}(d,e): their persistence preserves the steep fringe slopes required for high sensitivity even as the fringe amplitude drops, until the $4\phi$ oscillations are eventually washed out at $\mathcal{I} < 0.3$.

The dependence on single-photon purity $\mathrm{g^{(2)}(0)}$ follows a markedly different trend (\autoref{fig:paramsweeps}(d)). While the two-photon case exhibits limited variation for typical values of $\mathrm{g^{(2)}(0)} < 0.1$, the four-photon sensitivity is strongly degraded, approaching its maximum value only when $\mathrm{g^{(2)}(0)} < 10^{-3}$. In the ideal limit ($\mathrm{g^{(2)}(0)} \rightarrow 0$, $\mathcal{I} \rightarrow 1$), we find $S_{22 \rightarrow 31+13} \rightarrow \sqrt{3}$, reproducing the upper bound for a four-photon Holland-Burnett state \cite{Ma_2025}. We note that at large $\mathrm{g^{(2)}(0)}$, the single-photon input (red line) trends slightly above $S=1$. This reflects the presence of a non-negligible two-photon component in the input state rather than true super-sensitivity.

Taken together, our results indicate that multi-photon interference imposes significantly more stringent source requirements than schemes involving only two photons. Crucially, we find that achieving very low $\mathrm{g^{(2)}(0)}$ is the decisive factor for maximizing the contrast and phase sensitivity of four-photon interference. This stands in stark contrast to the two-photon case, where performance is primarily limited by photon indistinguishability ($\mathcal{I}$) for realistic source parameters.

\section*{Discussion}

Our experimental and theoretical results illustrate the fundamentally different structure and behaviour of four- and two-photon interference. The previously unresolved harmonic structure of four-photon fringes renders their maximum contrast and phase sensitivity surprisingly robust against photon distinguishability, but highly sensitive to multi-photon events ($\mathrm{g^{(2)}(0)}$). We propose that these are general features of HOM interference beyond the $N=2$ case. Defining generalised $N$-photon HOM interference as the simultaneous arrival of $N/2$ photons at each interferometer input port (where $N \in 2\mathbb{Z}^+$), the expectation value for the product of the photon numbers at the outputs is \cite{KIM199837}
\begin{equation}
\left< \hat{n_e} \hat{n_f} \right> = \frac{N}{8}(2N - (2+N) \mathrm{sin^2}(\phi)).
\label{eq:NphotonCoincExp}
\end{equation}
For $\phi = k \pi$ $(k \in \mathbb{Z})$, this reduces to $(N/2)^2$, indicating exactly $N/2$ photons at each output port and reproducing the input photon statistics. Meanwhile, generalising \autoref{eq:P22} to an $N$-photon Holland-Burnett state \cite{Xiang2013} shows that a $\ket{N-1,1}$ detection scheme produces fringes of the form:
\begin{equation}
P_{(N/2),(N/2) \rightarrow (N-1),1} = \frac{2^{-N}N!}{\left(\frac{N}{2}!\right)^2}(1-\mathrm{cos}(N\phi))/2,
\label{eq:NphotonFringe}
\end{equation}
which always exhibit minima with $P=0$ at $\phi = k \pi$. While these analytical expressions assume ideal Fock states, meaningful comparisons can still be made with our results. Modelling a four-photon input with non-unity indistinguishability (\autoref{fig:NPhotonFringes}(f)) confirms that $P_{2,2 \rightarrow 2,2} = 1$ (see supplementary material) and $P_{2,2 \rightarrow 3,1} = 0$ at the ``deep fringe'' positions ($\phi = k \pi$), matching the predictions of \autoref{eq:NphotonCoincExp} and \autoref{eq:NphotonFringe}. Therefore, we expect these interference minima at $\phi = k \pi$, which are insensitive to photon distinguishability, to occur in all equivalent ($N \in 2\mathbb{Z}^+$, $N/2$ photon inputs, $\ket{N-1,1}$ detection) interference fringes for $N \geq 4$. However, as the weighting of the pure state contributions in our model (see supplementary material) scale according to $\left(\mathrm{g^{(2)}(0)}\right)^N$, interference with larger $N$ will exhibit increasingly severe sensitivity to multi-photon emission.

To address this elevated sensitivity, reductions in $\mathrm{g^{(2)}(0)}$ could be achieved by shortening excitation pulses \cite{Liu2018,Hanschke2018} or employing 3-level excitation schemes, which have demonstrated $\mathrm{g^{(2)}(0)}<10^{-4}$ \cite{Hanschke2018,Schweickert2018}. For 3-level excitation, our tunable cavity enables the lifetime ratios required to simultaneously maintain high $\mathcal{I}$ \cite{PhysRevLett.125.233605}. Improvements in brightness are also desirable to reduce integration times and facilitate scaling to larger photon numbers. Increasing the source repetition rate ($1/\tau_{\rm rep}$) to $1/5T_1 = 3.3~\mathrm{GHz}$ would maintain $>99\%$ per-pulse emission probability \cite{Liu2018} while yielding a $> 40$-fold increase in photon flux. Currently, our end-to-end transmission in the four-photon configuration (excluding source and detector efficiencies) is $\sim 10 \%$, limited primarily by free-space to fibre coupling losses and extensive fibre path lengths. We estimate that optimizing mode matching, employing direct fibre splicing, and minimizing path lengths could increase this to $\sim50\%$. Due to the $\eta^N$ efficiency dependence of $N$-fold coincidences, this would boost four-photon rates by almost three orders of magnitude. Longer-term, integrated photonics platforms offer the potential for ultra-low loss implementations \cite{PhysRevLett.130.223601}.

Recent advances in photon-number-resolving (PNR) SNSPDs \cite{10.1063/5.0204340,Stasi2025} also offer a path to dramatically improve $N>2$ photon detection efficiency. We estimate our current three-photon detection probability at output $e$ is only $P_3 \simeq 0.05$; utilizing PNR SNSPDs with $>60\%$ three-photon efficiency \cite{Stasi2025} could boost our four-photon rates by an order of magnitude. Furthermore, PNR detectors allow simultaneous probing of multiple output states (e.g. $\ket{3_e 1_f}$ and $\ket{1_e 3_f}$) without the splitting losses associated with extra single-channel detectors, unlocking the increased phase sensitivity for metrology described in \autoref{fig:paramsweeps}(c,d).

Iterating the state preparation process -- through additional demultiplexing and HOM interference of $\ket{2}$ states to produce a $\ket{4,4}$ input -- could extend our scheme to higher photon numbers. Taking a source efficiency of $71\%$ \cite{Ding2025}, a repetition rate of $3.3~\mathrm{GHz}$, and detection efficiencies for 7 and 1 photons of 0.2 and 0.9 respectively \cite{Stasi2025}, we estimate an eight-photon coincidence rate of $10~\mathrm{kHz}$ (via $\ket{7,1}/\ket{1,7}$ detection) could be achieved with a realistic end-to-end transmission of $50\%$. Furthermore, we note the potential of QD-cavity systems in the strong coupling regime to produce larger $N$-photon states on-demand. Recent experiments have demonstrated degenerate photon pair emission \cite{Liu2025}, and theoretical studies suggest generation of up to 5-photon states is possible with realistic parameters \cite{PhysRevResearch.2.033489}. When operated with a high reflectivity top mirror \cite{Najer2019}, our QD open microcavity system can access the regimes required for such multi-photon state generation \cite{PhysRevResearch.2.033489}.

While this work focuses on NOON-like states, these represent just one class of multi-photon entangled states offering potential quantum advantage for phase estimation. Studies have shown that a single bright fringe of a Holland-Burnett state can outperform NOON states for phase estimation at $N>4$, with the advantage increasing with $N$ \cite{Xiang2013}. Other approaches have shown potential for sensitivity beyond the SQL by utilizing the photon-number correlations inherent to SPDC sources — - either by employing the full broad thermal distribution \cite{Matthews2016} or by heralding specific high-photon-number components \cite{Thekkadath2020}. Both strategies can exhibit higher loss tolerance than NOON states. Significant potential exists to exploit similar phenomena to realise optimised metrology schemes for QD sources. Notably, the coherent interaction between the resonant laser and the QD transition allows the preparation of coherent photon-number superpositions by tuning the pulse area away from $\Theta = \pi$ \cite{Loredo2019,Karli2024}. Combining these coherent photon number superpositions with the cascaded demultiplexing and interference approach introduced here offers a versatile toolbox for multi-photon quantum state engineering, offering extensive possibilities for future exploration of quantum metrology with novel multi-photon probes.

In summary, we have combined an efficient, on-demand QD SPS with active demultiplexing and HOM interference to experimentally realise direct quantum interference of up to four photons. High single-photon purity and indistinguishability enable high interference contrasts, allowing us to violate the standard quantum limit for phase sensitivity. Our experiments achieve order-of-magnitude rate increases over previous studies, resolving interference fringes with previously unseen harmonic structure and demonstrating a non-trivial relationship between indistinguishability and quantum interference beyond the two-photon case. An immediate consequence is that the maximum contrast of four-photon interference is highly sensitive to source multi-photon events ($\mathrm{g^{(2)}(0)>0}$) while remaining relatively insensitive to photon distinguishability ($\mathcal{I} < 1$) -- a situation almost the inverse of the two-photon case.

We believe that our results illustrate the vast potential of QD photon sources for multi-photon interference and optical quantum metrology experiments, with a clear path to increasing contrast, brightness and photon number. Beyond metrology, our results also hold significant relevance for photonic quantum computing and simulation. For example, Fock states of up to 5 photons have been demonstrated as a promising platform for quantum simulation \cite{Sturges2021}. Meanwhile, the harmonic structure of our multi-photon interference fringes offers new insights for linear optical quantum computing \cite{PhysRevLett.123.250503,Maring2024} and photonic indistinguishability purification schemes \cite{PhysRevLett.133.033604} -- protocols that inherently involve, but do not presently resolve, multi-photon interference.


\section*{Methods}

\subsection*{Experiment}

The QD SPS is driven optically by picosecond pulses (intensity full-width at half maximum between $3.6$ and $5.0~\mathrm{ps}$ \cite{Javadi_2023}) from a mode-locked Ti:Sapphire laser (Coherent Mira 900) operating at a repetition rate of $1/\tau_{\rm rep} = 76.3 ~\mathrm{MHz}$. The QD SPS comprises an open microcavity system with a dielectric top mirror and a semiconductor bottom mirror incorporated below the QD layer of the sample, as described in Ref. \cite{tomm_bright_2021}. The source is operated with the cavity length tuned such that it is resonant with the positive trion ($X^+$) state of a QD emitting at $\lambda \simeq 923~\mathrm{nm}$. The microcavity exhibits two linearly-polarised modes with a small splitting ($50~\mathrm{GHz}$), we operate in the ``blue collection'' configuration where the higher-frequency mode is resonant with the QD, and the laser excitation is through the lower-frequency mode \cite{Javadi_2023}. Scattered laser photons are rejected by a cross-polarized dark-field microscope configuration, achieving a measured signal-to-background ratio of $\sim 380:1$ (evaluated by turning off the QD emission by changing the applied bias voltage) when excited by a $\pi$-pulse.

Photons from the QD SPS are coupled to single mode fibre (Coherent 780-HP) to reach the state preparation apparatus. Here, a manual fibre polarization controller (Thorlabs FPC) aligns the photon polarisation with the crystal axes of a resonant polarization modulator (Qubig AM8-NIR). The modulator drive electronics are locked to the laser repetition rate, with the driver phase and amplitude tuned to achieve a maximum polarisation extinction ratio of around $20~\mathrm{dB}$. The modulator alternately routes subsequent photons to two spatial modes by rotating the photon polarization prior to a polarizing beamsplitter (PBS). A retroreflector (Newport UBBR1) mounted on a motorised linear stage (Zaber LSQ300) provides a variable optical delay on one of the two spatial modes.

The two free-space paths are then independently coupled into polarization maintaining (PM) fibre (Coherent PM780-HP), with the slow axis of each fibre aligned with the linear polarization axis of the free-space beams, achieving an in-fibre extinction ratio of $20~\mathrm{dB}$. For preparation of the $\ket{20}$ and $\ket{22}$ input states, the two spatial modes are interfered at a PM fibre beamsplitter (FiberLogix) with a near-ideal measured splitting ratio 0.511:0.489 at the QD source wavelength of $923~\mathrm{nm}$.

The two PM fibre paths are then coupled to the two input modes of a displaced Sagnac interferometer, with their slow axes aligned perpendicular to the top surface of the optical breadboard (Carbon Vision EC) on which the interferometer is built. Further fine-tuning of optical path length between the two inputs is achieved by a linear stage (Thorlabs XR50P) on which one of the two input fibre collimators is mounted.  The interferometer is built around a cube beamsplitter (Edmund Optics) with a measured splitting ratio of $0.505:0.495$ at the QD source wavelength, mounted on a precision tip-tilt-rotation mount (New Focus 9411-M). The Sagnac loop comprises three high reflectivity dielectric mirrors (Thorlabs BBE1-E03) with an anti-reflection coated glass plate (Thorlabs WG11050-AB) in each arm. One glass plate is tilted with respect to the optical axis by a precision piezo rotation stage (Newport CONEX-AG-PR100P) to produce the phase shift.

From the outputs of the interferometer, single mode fibre (Thorlabs 780HP) takes the photons to superconducting nanowire single-photon detectors (SNSPDs, Single Quantum EOS). The detectors have a single-photon detection efficiency of $\eta_{\rm PDE} = 0.82 \pm 0.05$ at the QD SPS wavelength \cite{tomm_bright_2021}. For the four-photon experiment, we split one of the two output fibres across 3 SNSPD channels using cascaded 75:25 and 50:50 single mode fibre beamsplitters (Thorlabs TW930R3A2 / TW930R3A5) to give approximately equal intensities at all 3 detectors. Detector count-rates and coincidence rates were measured using a time tagger system (Swabian Time Tagger Ultra Performance) connected to the SNSPD driver outputs.

\subsection*{Theoretical Model}
The theoretical model was developed using the input-output formalism. The output is described by linear transfer matrices for each of the optical elements in the interferometer: the beamsplitter, phase shift and optical losses. The result is then a set of transfer matrices that relate each of the input modes $a$ and $b$ to the output modes $e$ and $f$ (and the lossy mode $g$, if losses are considered). A full derivation of the model is presented in the supplementary material. 

To fit the model to our experimental data, a non-linear, least squares algorithm is used to optimise the fitting parameters: $\mathrm{g^2(0)}$, $\mathcal{I}$, $\eta_c$, $\eta_d$, $\eta_e$ and $\eta_f$. These parameters define the proportion of each pure state in the overall mixed input state. The fitting is conducted in Python using the ``LMfit'' package. To minimise mutual dependency and errors in the fitted parameters, the number of free parameters in any fit is minimised by fixing values that were previously measured independently (e.g. $\mathrm{g^2(0)}$) or determined in previous fits (e.g. $\eta_{c/d}$). As the various interference experiments are sensitive to different parameters, this approach allows us to determine a single consistent set of parameters that reproduces all our experiments with low parameter uncertainties. Full details on the fitting function, methodology and a table of extracted parameters and their uncertainties are presented in the supplementary material.

The phase sensitivity values presented in \autoref{fig:paramsweeps} were calculated by applying the equations presented in \cite{Okamoto_2008} to the results of our input-output model.

\section*{Acknowledgements}

This work was funded by the STFC (UK) grant ST/Y005031. A.J.B. acknowledges additional support from the EPSRC (UK) fellowship EP/W027909. R.J.W. acknowledges support from the Swiss National Science Foundation Project No. 200020\_204069 and the Innosuisse project SparQ. A.J.B. would also like to thank all the additional members of the Warburton group for their generous hospitality during his visits to Basel, in particular Giang Nam Nguyen and Yannik Fontana. L.B. would like to thank Pieter Kok and Jake Iles-Smith for helpful theoretical discussions.

\section*{Author Contributions}

A.J.B. and R.J.W. conceived and directed the project. A.J.B., L.B. and C.L.P. designed, assembled and aligned the displaced Sagnac interferometer. S.R.V. and A.L. fabricated and processed the semiconductor device. M.R.H., M.A.M. and T.B. aligned, optimised and operated the QD-microcavity single-photon source during the experiments. A.J.B., L.B. and M.R.H. performed the multi-photon interference experiments with support from M.A.M. and T.L.B.. L.B. performed the input-output formalism modelling of the system and the Fisher information calculations. A.J.B. and L.B. wrote the manuscript with input from all authors.


\bibliography{bibliography}

\end{document}


\title[Article Title]{Supplementary Material: Direct Four-Photon Interference with a High-Efficiency Quantum Dot Source}

\author*[1]{\fnm{Alistair J.} \sur{Brash}}\email{a.brash@sheffield.ac.uk}

\author[1]{\fnm{Luke} \sur{Brunswick}}

\author[2]{\fnm{Mark R.} \sur{Hogg}}

\author[1]{\fnm{Catherine L.} \sur{Phillips}}

\author[2]{\fnm{Malwina A.} \sur{Marczak}}

\author[2]{\fnm{Timon} \sur{Baltisberger}}

\author[3]{\fnm{Sascha R.} \sur{Valentin}}

\author[3]{\fnm{Arne} \sur{Ludwig}}

\author[2]{\fnm{Richard J.} \sur{Warburton}}

\affil*[1]{\orgdiv{School of Mathematical and Physical Sciences}, \orgname{University of Sheffield}, \orgaddress{\street{Hounsfield Road}, \city{Sheffield}, \postcode{S3 7RH}, \country{United Kingdom}}}

\affil[2]{\orgdiv{Department of Physics}, \orgname{University of Basel}, \orgaddress{\street{Klingelbergstrasse 80}, \city{Basel}, \postcode{4056}, \country{Switzerland}}}

\affil[3]{\orgdiv{Faculty of Physics and Astronomy}, \orgname{Ruhr-Universit\"{a}t Bochum}, \orgaddress{\street{Universit\"{a}tsstrasse 150}, \city{Bochum}, \postcode{44801}, \country{Germany}}}

\maketitle

\section{Input-Output Formalism}

Whilst the analytical expressions shown in the main text were relatively easily derived for an ideal system, this is not the case when considering realistic imperfections such as non-unity single-photon purity ($\mathrm{g^{(2)}(0) \neq 0}$). To address this, we develop a model using the input-output formalism, using the transfer matrix approach.

A schematic of the theoretical model is shown in \autoref{fig:schematic}(a) of the main manuscript. Photons are coupled into the system in one of two input modes \textit{a} or \textit{b}; these modes are then subjected to a series of matrix operations -- representing the optical elements and losses in the experimental setup -- before the detection probabilities are measured in the output modes \textit{e} and \textit{f}. Each matrix relates the input and output states by the linear transfer relation
\begin{equation}
     \begin{pmatrix}
         r_{i,out}\\r_{j,out}
     \end{pmatrix} =  M\begin{pmatrix}
         r_{i,in}\\r_{j,in}
     \end{pmatrix},\label{equ:lineartransfer}
\end{equation}where $r_{i,in}$ and $r_{i,out}$ are the input and output mode operators respectively, and \textit{M} is the transfer matrix of the optical element. For a dielectric beamsplitter with a 50:50 splitting ratio, the matrix element is defined as
\begin{equation}
    B = \frac{1}{\sqrt{2}} \begin{pmatrix}
        1 & i\\
        i & 1
\end{pmatrix}.\label{equ:5050BS}
\end{equation} The form of the phase shift matrix is
\begin{equation}
    S = \begin{pmatrix}
        1 & 0\\
        0 & e^{i\phi}
\end{pmatrix},\label{equ:phaseshiftmatrix}
\end{equation}where $\phi$ is the phase shift imparted between the two interferometer modes ($c$/$d$). In the case where photon loss is considered, either inside the interferometer or due to the imperfect detection efficiency, each mode $i$ is considered to couple independently to a lossy mode $g$ with a probability $\eta_i$, $i\in\{c,d,e,f \}$. The input-output relation for a single mode in the interferometer is given by
\begin{equation}
     \begin{pmatrix}
         a_{i,out}\\g
     \end{pmatrix} = \begin{pmatrix}
        \eta_i & 1-\eta_i\\
        1-\eta_i & \eta_i
\end{pmatrix} \begin{pmatrix}
         a_{i,in}\\g
     \end{pmatrix}.\label{equ:lossmatrix}
\end{equation}The detection probability for a particular output state, given a particular input state, is defined as the sum over all possible outputs which satisfy the detection conditions (at least $e$ photons in mode $e$ and $f$ photons in mode $f$).
\begin{equation}
    P_{I \rightarrow ef} = \sum_{i = e}^{T-f} \sum_{j = f}^{T-i} \left| p_{I \rightarrow ij} \right|^{2},
    \label{equ:measurementprobability}
\end{equation}where $I$ is the input pure state at modes $a$ and $b$, $T$ is the total number of photons input into the system, $e$ and $f$ are the minimum number of photons required to be detected in modes $e$ and $f$ respectively, and $\left| p_{I \rightarrow ij} \right|^{2}$ is the probability of detecting an output state with $i$ photons in mode $e$ and $j$ photons in mode $f$ from a given input state $I$. The minimum values for $i/j$ are bounded to the photon number requirements ($e/f$) ensuring that we sum over only states with at least the required photon number in each mode. This is important as we consider both the presence of additional (distinguishable) photons in the input state due to multi-photon emission events, leading to situations where $T > e+f$. 

\autoref{equ:measurementprobability} gives the probability of measuring a specific photon number state at the output of the system, given a specific pure input state. To consider the impact of a mixed input state, the weighted probabilities from each input pure state are summed:
\begin{equation}
    P^{total}_{ab \rightarrow ef} = \sum_{ L } \alpha_{L}P_{L \rightarrow ef},
\label{equ:totalweightedprobability}
\end{equation}where $L \in \{I_1,I_2...I_n\}$ is the set of input pure state permutations considered for each measurement case, these permutations cover the possible multi-photon, and distinguishable photon emission cases for the target input state $a,b$. $\alpha_{L}$ is the weighted probability of each input pure state such that $\sum_{I} \alpha_{I} = 1$. The values of $\alpha_L$ are related to $\mathrm{g^{(2)}(0)}$ and $\mathcal{I}$. $P_{I \rightarrow ef}$, obtained from \autoref{equ:measurementprobability}, is the probability of measuring a specific photon number state ($e,f$) at the output of the system, for a given input pure state $L$. We truncate our photon number basis at $N = 5$ as we work in a regime where the source multi-photon emission probability $\mathrm{g^{(2)}(0) \ll 1}$, meaning that the probability of events involving $> 5$ photons is vanishingly small.

\section{Derivation of Coefficients From Input-Output Model}
This section will detail the derivation of the pure state outputs from our input-output model which are combined in a weighted sum to produce the mixed input state result as detailed above. We show the generalised derivation for each input mode $a$ and $b$ as the result applies to any input state.

Firstly, we consider the input modes $a$ and $b$, which we can denote as the column vector $     \begin{pmatrix}
         a^\dagger\\b^\dagger
         \end{pmatrix}$.
Using the schematic in \autoref{fig:schematic}(a) we see that the first transfer matrix to be applied to the input state is the beamsplitter. In our model, we assume that our beamsplitter has a 50:50 splitting ratio as the measured splitting ratio (50.5:49.5) does not significantly deviate from this.
Applying \autoref{equ:lineartransfer}, and substituting $M$ for the matrix in \autoref{equ:5050BS} the output state from the beamsplitter is given by:
\begin{equation}
        \begin{pmatrix}
            c^\dagger\\d^\dagger
        \end{pmatrix} = \frac{1}{\sqrt{2}}\begin{pmatrix}
        1 & i\\
        i & 1
\end{pmatrix}\begin{pmatrix}
    a^\dagger\\b^\dagger
\end{pmatrix}. 
\label{equ:firstBSpass}
\end{equation}
From \autoref{equ:firstBSpass}, it follows that:
\begin{equation}
    a^\dagger \xrightarrow{} \frac{1}{\sqrt{2}}(c^\dagger + id^\dagger),
    \label{equ:aafterBS}
\end{equation}
\begin{equation}
    b^\dagger \xrightarrow{} \frac{1}{\sqrt{2}}(ic^\dagger + d^\dagger).
    \label{equ:bafterBS}
\end{equation}
\\
Next, we apply the phase shift matrix from \autoref{equ:phaseshiftmatrix}:
\begin{equation}
        \begin{pmatrix}
            c^\dagger_{out}\\d^\dagger_{out}
        \end{pmatrix} = \begin{pmatrix}
        1 & 0\\
        0 & e^{i\phi}
\end{pmatrix}\begin{pmatrix}
    c^\dagger_{in}\\d^\dagger_{in}
\end{pmatrix}. \label{equ:phaseshiftoutput}
\end{equation}
This transforms the states \autoref{equ:aafterBS} and \autoref{equ:bafterBS} to:
\begin{equation}
    \frac{1}{\sqrt{2}}(c^\dagger + id^\dagger) \xrightarrow{} \frac{1}{\sqrt{2}}(c^\dagger + ie^{i\phi}d^\dagger),
    \label{equ:aafterPS}
\end{equation}
\begin{equation}
    \frac{1}{\sqrt{2}}(ic^\dagger + d^\dagger) \xrightarrow{} \frac{1}{\sqrt{2}}(ic^\dagger + e^{i\phi}d^\dagger).
    \label{equ:bafterPS}
\end{equation}
\\
We then consider the effect of loss within the interferometer using the transfer matrix of the form shown in \autoref{equ:lossmatrix}. We apply one matrix to each mode $c$ and $d$ as follows:
\begin{equation}
            \begin{pmatrix}
            c^\dagger_{out}\\g^\dagger_{out}
        \end{pmatrix} = \begin{pmatrix}
        \eta_c & 1-\eta_c\\
        1-\eta_c & \eta_c
\end{pmatrix}\begin{pmatrix}
    c^\dagger_{in}\\g^\dagger_{in}
\end{pmatrix}, \label{equ:LILetac}
\end{equation}
\\
\begin{equation}
            \begin{pmatrix}
            d^\dagger_{out}\\g^\dagger_{out}
        \end{pmatrix} = \begin{pmatrix}
        \eta_d & 1-\eta_d\\
        1-\eta_d & \eta_d
\end{pmatrix}\begin{pmatrix}
    d^\dagger_{in}\\g^\dagger_{in}
\end{pmatrix}, \label{equ:LILetad}
\end{equation}
\\
leading to the transformation of the input states:
\begin{equation}
        \frac{1}{\sqrt{2}}(c^\dagger + ie^{i\phi}d^\dagger) \xrightarrow{} \frac{1}{\sqrt{2}}(\eta_cc^\dagger + i\eta_de^{i\phi}d^\dagger + ((1-\eta_c) + ie^{i\phi}(1-\eta_d))g^\dagger),
    \label{equ:aafterLIL}
\end{equation}
\begin{equation}
        \frac{1}{\sqrt{2}}(ic^\dagger + e^{i\phi}d^\dagger) \xrightarrow{} \frac{1}{\sqrt{2}}(i\eta_cc^\dagger + \eta_de^{i\phi}d^\dagger + (i(1-\eta_c) + e^{i\phi}(1-\eta_d))g^\dagger),
    \label{equ:bafterLIL}
\end{equation}\\
as we assume no photons are input in mode $g$.\\
Next, the second pass of the beamsplitter is evaluated, transfering the photons into the output modes $e$ and $f$:
\begin{equation}
           \begin{pmatrix}
            e^\dagger\\f^\dagger
        \end{pmatrix} = \frac{1}{\sqrt{2}}\begin{pmatrix}
        1 & i\\
        i & 1
\end{pmatrix}\begin{pmatrix}
    c^\dagger\\d^\dagger
\end{pmatrix}.
\label{equ:2ndBSpass}
\end{equation}
Similarly to \autoref{equ:aafterBS} and \autoref{equ:bafterBS}, the modes evolve as such:
\begin{equation}
    c^\dagger \xrightarrow{} \frac{1}{\sqrt{2}}(e^\dagger + if^\dagger),
    \label{equ:ctoef}
\end{equation}
\begin{equation}
    d^\dagger \xrightarrow{} \frac{1}{\sqrt{2}}(ie^\dagger + f^\dagger).
    \label{equ:dtoef}
\end{equation}
We note that the beamsplitter has no effect on the photons in the lossy mode $g$ as these photons are considered to have left the system.
\\
The evolution of the input states is then:
\begin{multline}
    \frac{1}{\sqrt{2}}(\eta_cc^\dagger + i\eta_de^{i\phi}d^\dagger + ((1-\eta_c) + ie^{i\phi}(1-\eta_d))g^\dagger)\\
         \xrightarrow{}
         \frac{1}{\sqrt{2}}\left(\left(\frac{\eta_c}{\sqrt{2}}-\frac{\eta_d}{\sqrt{2}}e^{i\phi}\right)e^\dagger +\left( i\frac{\eta_c}{\sqrt{2}} + i\frac{\eta_d}{\sqrt{2}}e^{i\phi} \right)f^\dagger + ((1-\eta_c) + ie^{i\phi}(1-\eta_d))g^\dagger\right),
    \label{equ:aafter2ndBS}
\end{multline}
\begin{multline}    
         \frac{1}{\sqrt{2}}(i\eta_cc^\dagger + \eta_de^{i\phi}d^\dagger + (i(1-\eta_c) + e^{i\phi}(1-\eta_d))g^\dagger)\\ 
         \xrightarrow{} 
         \frac{1}{\sqrt{2}}\left(\left(i\frac{\eta_c}{\sqrt{2}}+i\frac{\eta_d}{\sqrt{2}}e^{i\phi}\right)e^\dagger +\left( \frac{\eta_d}{\sqrt{2}}e^{i\phi} - \frac{\eta_c}{\sqrt{2}} \right)f^\dagger + (i(1-\eta_c) + e^{i\phi}(1-\eta_d))g^\dagger\right).
    \label{equ:bafter2ndBS}
\end{multline}

Finally, the effect of detector efficiency (modelled as loss in modes $e$ and $f$) is evaluated using the same approach as \autoref{equ:LILetac} and \autoref{equ:LILetad}:
\begin{equation}
                \begin{pmatrix}
            e^\dagger_{out}\\g^\dagger_{out}
        \end{pmatrix} = \begin{pmatrix}
        \eta_e & 1-\eta_e\\
        1-\eta_e & \eta_e
\end{pmatrix}\begin{pmatrix}
    e^\dagger_{in}\\g^\dagger_{in}
\end{pmatrix}, \label{equ:LDetae}
\end{equation}
\begin{equation}
                \begin{pmatrix}
            f^\dagger_{out}\\g^\dagger_{out}
        \end{pmatrix} = \begin{pmatrix}
        \eta_f & 1-\eta_f\\
        1-\eta_f & \eta_f
\end{pmatrix}\begin{pmatrix}
    f^\dagger_{in}\\g^\dagger_{in}
\end{pmatrix}. \label{equ:LDetaf}
\end{equation}\\
The final input-output states are then given by:
\begin{equation}
    a^\dagger \xrightarrow{} \frac{1}{\sqrt{2}} \left(Ae^\dagger + Bf^\dagger + Cg^\dagger \right),\label{equ:finalaevo}
\end{equation}
\begin{equation}
    b^\dagger \xrightarrow{} \frac{1}{\sqrt{2}} \left(De^\dagger + Ef^\dagger + Fg^\dagger \right),\label{equ:finalbevo}
\end{equation}
where:
\begin{equation*}
    \begin{split}
    A& =\eta_e\left(\frac{\eta_c}{\sqrt{2}}-\frac{\eta_d}{\sqrt{2}}e^{i\phi}\right), \\
    B& = \eta_f\left( i\frac{\eta_c}{\sqrt{2}} + i\frac{\eta_d}{\sqrt{2}}e^{i\phi} \right),\\
    C& = (1-\eta_c) + ie^{i\phi}(1-\eta_d) + (1-\eta_e)\left(\frac{\eta_c}{\sqrt{2}}-\frac{\eta_d}{\sqrt{2}}e^{i\phi}\right) + (1-\eta_f)\left( i\frac{\eta_c}{\sqrt{2}} + i\frac{\eta_d}{\sqrt{2}}e^{i\phi} \right), \\
    D& = \eta_e\left(i\frac{\eta_c}{\sqrt{2}}+i\frac{\eta_d}{\sqrt{2}}e^{i\phi}\right),\\
    E& = \eta_f \left(\frac{\eta_d}{\sqrt{2}}e^{i\phi} - \frac{\eta_c}{\sqrt{2}} \right), \\
    F& = i(1-\eta_c) + e^{i\phi}(1-\eta_d) + (1-\eta_e)\left(i\frac{\eta_c}{\sqrt{2}}+i\frac{\eta_d}{\sqrt{2}}e^{i\phi}\right) + (1-\eta_f) \left(\frac{\eta_d}{\sqrt{2}}e^{i\phi} - \frac{\eta_c}{\sqrt{2}} \right).\\ 
\end{split}
\end{equation*}\\
\autoref{equ:finalaevo} and \autoref{equ:finalbevo} give the input-output relationship for a single-photon input into mode $a$ or $b$, respectively. These expressions can then be used to calculate the output state of a given input pure state. From this state, the measurement probability of specific output states can be calculated (see \autoref{equ:measurementprobability}). \\
As an example, here we derive the expression for the probability of measuring the output state $\ket{3_e1_f}$ with an input state $\ket{2_a1_{a'}2_b}$. This state represents the case where our four-photon state preparation is successful in creating two indistinguishable, two-photon Fock states in modes $a$ and $b$, respectively, whilst an additional distinguishable photon is present in mode $a$ due to a multi-photon excitation event during the state preparation. We use $'$ to denote which photons are distinguishable from one another.\\
The input state can be written in terms of the mode operators and the vacuum state $\ket{0}$:
\begin{equation}
    \ket{2_a1_{a'}2_b}_{in} = \frac{a^{\dagger2}}{\sqrt{2}}\frac{b^{\dagger2}}{\sqrt{2}} a'^\dagger\ket{0}.
    \label{equ:212input}
\end{equation}\\The factors of $\frac{1}{\sqrt{2}}$ come from the required normalisation for $N$ indistinguishable emitters in the same mode:
\begin{equation}
    \ket{N_m}=\frac{1}{\sqrt{N!}}\left(a_m^\dagger\right)^N \ket{0}.
    \label{equ:fockstateNorm}
\end{equation}\\
Using \autoref{equ:finalaevo} and \autoref{equ:finalbevo}, the output state is:
\begin{multline}
        \ket{2_a1_{a'}2_b}_{out} = \frac{1}{2} \times \left(\frac{1}{\sqrt{2}}\left(Ae^\dagger + Bf^\dagger + Cg^\dagger \right)\right)^2 \times \left(\frac{1}{\sqrt{2}}\left(De^\dagger + Ef^\dagger + Fg^\dagger \right)\right)^2 \\\times \left(\frac{1}{\sqrt{2}}\left(Ae'^\dagger + Bf'^\dagger + Cg'^\dagger \right)\right)\ket{0}.
    \label{equ:212fulloutput}
\end{multline}\\
To find the probability of measuring the output $\ket{3_e1_f}$, \autoref{equ:measurementprobability} tells us we must sum over the probabilities of all the possible output permutations which fulfil the photon number requirement. Therefore we must calculate the coefficients of each output state $\ket{m,n}$, with $m\geq 3$ and $n \geq 1$ simultaneously. In our example, the relevant output permutations are: $\ket{3_e1_{e'}1_f},\ket{3_e1_f1_{f'}},\ket{2_e1_{e'}2_f},\ket{4_e1_{f'}},\ket{3_e1_{f}1_{g'}},\ket{3_e1_{f'}1_{g}},\ket{2_e1_{e'}1_{f}1_{g}}$.\\

After summing over the probabilities calculated for each of these permutations, the final expression is:
\begin{multline}
        P_{3,1} = \left(\frac{1}{8\sqrt{2}}\right)^2\biggl[ 24\left|A^2BD^2 \right|^2 +4\left|A\left( A^2E^2 + B^2D^2 + 4ABDE \right) \right|^2\\ + 6\left|A\left( 2A^2DE +2ABD^2 \right) \right|^2 + 6\left|B\left( 2A^2DE + 2ABD^2 \right) \right|^2 + 6\left|C\left( 2A^2DE + 2ABD^2 \right) \right|^2 \\+ 
        6\left|B\left( 2A^2DF + 2ACD^2 \right) \right|^2 +2\left|A\left( 2A^2EF + 4ABDF + 4ACDE + 2BCD^2 \right) \right|^2 \biggr].
\end{multline}\\
This expression is then combined with the other possible input state permutations in the weighted sum from \autoref{equ:totalweightedprobability} to produce the modelled interference fringes. The weighting of each input state is discussed in detail in \autoref{sec:fittingweights}
\section{Fitting Function}
To accurately fit our model to the experimental data, we utilised a non-linear, least squares algorithm to optimise the fitting parameters. The fitting was conducted in Python using the "LMfit" module.

\subsection{Weighting of Input States}
\label{sec:fittingweights}
As shown in \autoref{equ:totalweightedprobability} and discussed above, we create our mixed input state by summing over the output probabilities for a set of input pure states. Each pure state is multiplied by a weighting, $\alpha_L$, which corresponds to the proportion of the mixed state described by each pure state.\\
The set of pure states $L$ consists of all possible states (up to $N = 5$ for the four-photon interference fringes) which could be produced by a realistic photon source using the state preparation procedures outlined in \autoref{fig:schematic} for each of the four ideal cases: $\ket{1_a0_b},\ket{1_a1_b},\ket{2_a0_b}$ and $\ket{2_a2_b}$. As for a realistic single-photon source $\mathrm{g^{(2)}(0)}\neq 0$ and $\mathcal{I} \neq1$, it follows that $L$ must consist of pure states with additional photons, and with distinguishable photons. The value of each weighting should therefore be proportional to the values of $\mathrm{g^{(2)}(0)}$ and $\mathcal{I}$.\\
\subsubsection{Single-Photon Input Weighting}
The single-photon interference case is the most simple as only two pure states are considered: the ideal case $\ket{1_a0_b}$ and the case where an extra, distinguishable photon is produced $\ket{1_a1_{a'}0_b}$. Here, the weighting of each state is proportional only to the value of $\mathrm{g^{(2)}(0)}$, as we assume any photons present due to the non-zero value of $\mathrm{g^{(2)}(0)}$ are fully distinguishable from the stream of intentionally created single-photons:
\begin{equation*}
\begin{split}
    \alpha_{1_a0_b} &= 1-\mathrm{g^{(2)}(0)},\\
    \alpha_{1_a1_{a'}0_b}& = \mathrm{g^{(2)}(0)}.
\end{split}   
\end{equation*}
\subsubsection{Two-Photon Input Weighting}
For the two-photon input cases, the weighting coefficients become dependent on both $\mathcal{I}$ and $\mathrm{g^{(2)}(0)}$ as there is now a chance for the input photons to be either distinguishable or indistinguishable. As we only consider the indistinguishability of the intended input photon state (i.e. not the photons present due to $\mathrm{g^{(2)}(0)} \neq 0$), the weightings are linearly proportional to $\mathcal{I}$. However, as there are now two excitations of our photon source there is now an additional chance to produce an extra photon in the state preparation, compared to the single-photon case. Therefore, the weightings have a quadratic dependence on $\mathrm{g^{(2)}(0)}$.\\
The weightings for the $\ket{1_a1_b}$ input are:
\begin{equation*}
\begin{split}
    \alpha_{1_a1_b} &= \mathcal{I}\times\left(1-\mathrm{g^{(2)}(0)}\right)^2, \\
    \alpha_{1_a1_{b'}} &=\left(1-\mathcal{I}\right)\times\left(1-\mathrm{g^{(2)}(0)}\right)^2, \\
    \alpha_{1_a1_{a'}1_b}& = \mathcal{I}\times \mathrm{g^{(2)}(0)}\times \left( 1-\mathrm{g^{(2)}(0)} \right),\\
    \alpha_{1_a1_b1_{b'}}& = \mathcal{I}\times \mathrm{g^{(2)}(0)}\times \left( 1-\mathrm{g^{(2)}(0)} \right),\\
    \alpha_{1_a1_{a'}1_b1_{b''}}& =\mathcal{I}\times \left( \mathrm{g^{(2)}(0)} \right)^2, \\
    \alpha_{1_a1_{a'}1_{b''}} &= \left( 1-\mathcal{I} \right)\times \mathrm{g^{(2)}(0)} \times \left( 1-\mathrm{g^{(2)}(0)}\right), \\
    \alpha_{1_a1_{b'}1_{b''}} &= \left( 1-\mathcal{I} \right)\times \mathrm{g^{(2)}(0)} \times \left( 1-\mathrm{g^{(2)}(0)}\right), \\
    \alpha_{1_a1_{a'}1_{b''}1_{b'''}} &= \left( 1-\mathcal{I} \right)\times \left(\mathrm{g^{(2)}(0)}\right)^2.\\
\end{split}   
\end{equation*}

The weightings for the $\ket{2_a0_b}$ input are:
\begin{equation*}
    \begin{split}
            \alpha_{2_a0_b} &= \mathcal{I}\times\left(1-\mathrm{g^{(2)}(0)}\right)^2, \\
    \alpha_{1_a1_{a'}0_b} &=\frac{1}{2}\times\left(1-\mathcal{I}\right)\times\left(1-\mathrm{g^{(2)}(0)}\right)^2, \\
    \alpha_{2_a1_{a'}0_b} &= 2\times\mathcal{I}\times \mathrm{g^{(2)}(0)}\times \left( 1-\mathrm{g^{(2)}(0)} \right), \\
        \alpha_{2_a1_{a'}1_{a''}0_b} &= \mathcal{I}\times \left(\mathrm{g^{(2)}(0)}\right)^2, \\
            \alpha_{1_a1_{a'}1_{a''}0_b} &= \left(1-\mathcal{I}\right)\times \mathrm{g^{(2)}(0)}\times \left( 1-\mathrm{g^{(2)}(0)} \right), \\
        \alpha_{1_a1_{a'}1_{a''}1_{a'''}0_b} &= \frac{1}{2}\times\left(1-\mathcal{I}\right)\times \left(\mathrm{g^{(2)}(0)}\right)^2. \\
    \end{split}
\end{equation*} Where the factors of $\frac{1}{2}$ come from the reduced state preparation success probability for two distinguishable photons and the factor of $2$ comes from the two permutations of the $\alpha_{2_a1_{a'}0_b}$ state.
\subsection{Four-Photon Weighting}
Finally, in the four-photon case, the complexity of the weighting increases again. We must account not only for the characteristics of the photon source, but also our state preparation procedure. As we rely on probabilistic quantum interference to route the indistinguishable photon pairs into the correct temporal and spatial modes, we must account for the case where this quantum interference fails. This introduces factors of $\frac{1}{2}$ and $\frac{1}{4}$ to weightings where one or both of these quantum interference steps fails. We assume that if both quantum interference steps are successful, that these quantum states are indistinguishable from one another. This gives our weightings a quadratic dependence on $\mathcal{I}$. As there are four single-photon excitation events in the state preparation, the weightings are have a quartic dependence on $\mathrm{g^{(2)}(0)}$. As discussed previously, the photon number basis is truncated to $N = 5$, meaning our model is only valid for values of $\mathrm{g^{(2)}(0)} \leq 0.1$ as at larger values of $\mathrm{g^{(2)}(0)}$ the effect of $N > 5$ photon states becomes increasingly significant. As shown in \autoref{fig:G2andTdep}, the $\mathrm{g^{(2)}(0)}$ of our source when obtaining the interference fringe data is well below this threshold. The weightings for the considered pure states are:
\begin{equation*}
    \begin{split}
        \alpha_{2_a2_b} &= \mathcal{I}^2 \times \left( 1-\mathrm{g^{(2)}(0)} \right)^4,\\
        \alpha_{1_a1_{a'}1_{b''}1_{b'''}} &=\frac{1}{4}\times\left(1-\mathcal{I}\right)^2 \times \left( 1-\mathrm{g^{(2)}(0)} \right)^4, \\
        \alpha_{2_a1_{a'}2_b} &=2\times\mathcal{I}^2\times \mathrm{g^{(2)}(0)}\times\left(1-\mathrm{g^{(2)}(0)}\right)^3, \\
        \alpha_{2_a2_b1_{b'}} &=2\times\mathcal{I}^2\times \mathrm{g^{(2)}(0)}\times\left(1-\mathrm{g^{(2)}(0)}\right)^3, \\
        \alpha_{2_a1_{a'}2_b1_{b''}} &=4\times\mathcal{I}^2\times \left(\mathrm{g^{(2)}(0)}\right)^2\times\left(1-\mathrm{g^{(2)}(0)}\right)^2, \\
        \alpha_{1_a1_{a'}1_{a''}1_{b'''}1_{b''''}} &=\frac{1}{2}\times\left(1-\mathcal{I}\right)^2\times \mathrm{g^{(2)}(0)}\times\left(1-\mathrm{g^{(2)}(0)}\right)^3, \\
        \alpha_{1_a1_{a'}1_{b''}1_{b'''}1_{b''''}} &=\frac{1}{2}\times\left(1-\mathcal{I}\right)^2\times \mathrm{g^{(2)}(0)}\times\left(1-\mathrm{g^{(2)}(0)}\right)^3,  \\
        \alpha_{1_a1_{a'}2_b} &= \left(1-\mathcal{I}\right)\times\mathcal{I}^2\times\left(1-\mathrm{g^{(2)}(0)}\right)^4,\\
        \alpha_{2_a1_b1_{b'}} &= \left(1-\mathcal{I}\right)\times\mathcal{I}^2\times\left(1-\mathrm{g^{(2)}(0)}\right)^4, \\
        \alpha_{1_a1_{a'}1_b1_{b'}} &= \left(1-\mathcal{I}\right)^2\times\mathcal{I}\times\left(1-\mathrm{g^{(2)}(0)}\right)^4.           
    \end{split}
\end{equation*}

\subsection{Fitting Parameters}
The fitting function takes 6 physical parameters ($\mathrm{g^{(2)}(0)}$, $\mathcal{I}$, $\eta_c$, $\eta_d$, $\eta_e$ and $\eta_f$). Due to mutual dependencies between the parameters present in the fitting function, fitting the experimental data with all the variables as free parameters returns extremely large uncertainties. To avoid this, we fix several of the parameters, and introduce sensible limits to those remaining, to reduce the parameter space available to the fitting function. We also exploit the fact that all measurements were taken sequentially during a single experimental run, ensuring consistent conditions and allowing us to fix some values extracted from single- and two-photon fits in the four-photon fit.

Initially, we consider the $\ket{1_{a},0_{b}}$ input case, for which the fringes are unaffected by the values of $\mathcal{I}$ and $\eta_f$.
\begin{figure}
    \centering
    \includegraphics[width=1\linewidth]{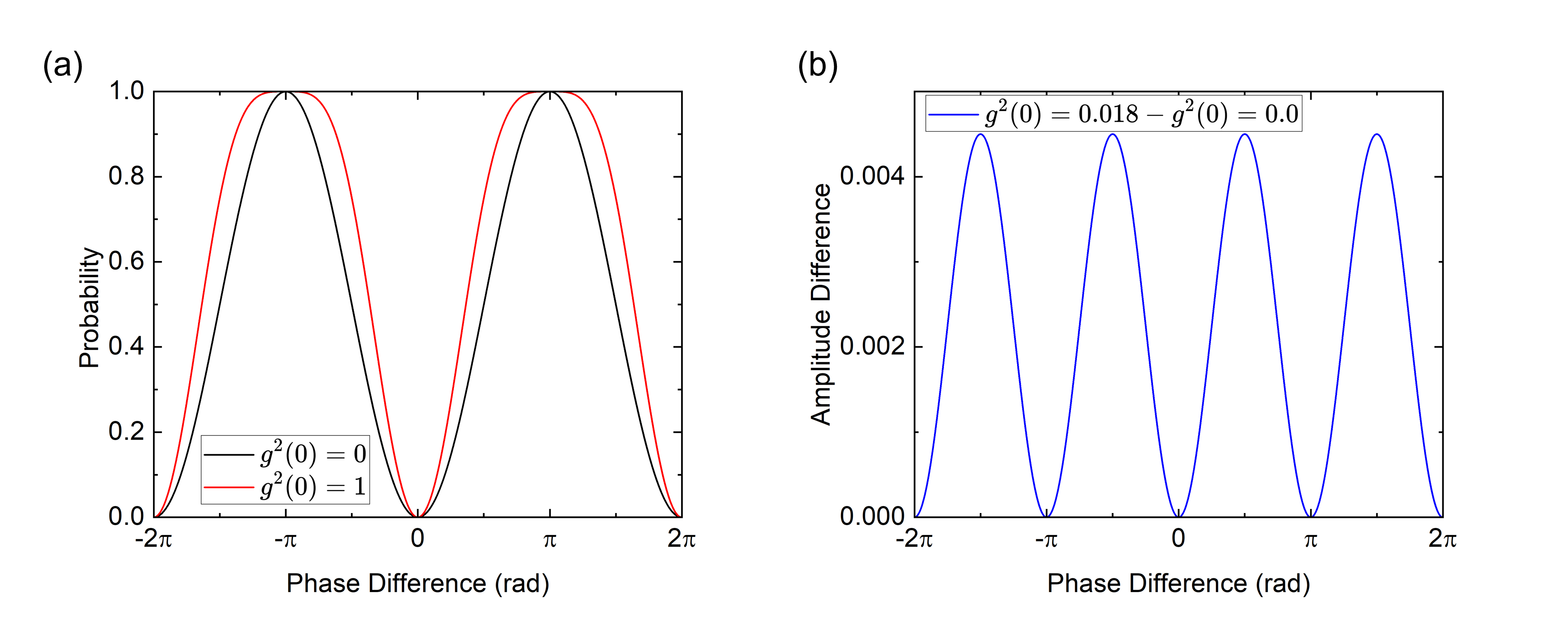}
    \caption{(a) Comparison between the single-photon fringes obtained from the input-output model for $\mathrm{g^{(2)}(0)} = 0$ and $\mathrm{g^{(2)}(0)} = 1$ illustrating the deviation from the perfectly sinusoidal relationship between phase difference and detection probability when multi-photon events are present. (b) Difference between the single-photon fringes for the ideal case: $\mathrm{g^{(2)}(0)} = 0$ and the realistic source case: $\mathrm{g^{(2)}(0)} = 0.018$.}
    \label{fig:singlephotonfringecomp}
\end{figure}
As discussed in the main text - and later in detail in \autoref{sec:phasecalib} - due to the finite surface flatness of the glass plate used to impart the phase shift between modes $c$ and $d$, an unknown relationship between the phase plate angle and phase shift is present. To counteract this, we calibrate the relationship between the phase plate angle and imparted phase shift by assuming the single-photon fringes should follow a perfectly sinusoidal relationship between the phase shift and fringe amplitude. However, \autoref{fig:singlephotonfringecomp} shows that the presence of multi-photon events also slightly alters the shape of the single-photon fringe pattern. This alteration will be subsumed into the phase calibration outlined in \autoref{sec:phasecalib}.  While this change is small for the value of $\mathrm{g^{(2)}(0)}$ measured for our QD source, it is still likely to impact the result of the single-photon fringe fit, removing the impact of $\mathrm{g^{(2)}(0)}$ from the data. Indeed, we find that the best fit for the single-photon data occurs when $\mathrm{g^{(2)}(0)} = 0$. Thus, when fitting the single-photon fringes, we fix $\mathrm{g^{(2)}(0)} = 0$. Further to this, we fix $\eta_e = 1$ as the shape of the fringes is only affected by the value of $\eta_e$ when $\mathrm{g^{(2)}(0)} \neq 0$. Finally, we fix the value of $\eta_c = 0.8034$, as there is a strong mutual dependence between $\eta_c$ and $\eta_d$ due to the symmetry of the interferometer. The value $\eta_c = 0.8034$ is calculated from the measured efficiencies of the optical components used to build the Sagnac loop: two passes of the beamsplitter ($0.91^2$), three mirror reflections ($0.9913^3$) and transmission through the glass plate once ($0.996$).

For the $\ket{2_a,0_b}$ input case, we must fix both $\mathrm{g^{(2)}(0)}$ and $\mathcal{I}$ as the input states produced by these parameters share the same phase relationship as the indistinguishable case, albeit with a different amplitude. Therefore, it is not possible to fit these parameters from the $\ket{2_a, 0_b}$ fringes. We choose to fix $\mathrm{g^{(2)}(0)} = 0.018$  from our independent measurement of the source in \autoref{fig:G2andTdep} of the main manuscript, and $\mathcal{I} = 1.00$ (presented as N/A in \autoref{tab:fitvalues} as the true value is undetermined by this fit). We are no longer required to fix $\mathrm{g^{(2)}(0)} = 0$ as the influence of $\mathrm{g^{(2)}(0)}$ on the fringes is no longer removed by the phase calibration. We also choose to fix $\eta_c$ to the previously calculated value. We note that the amplitude of the oscillations in the heights of the fringes is related to the ratio of $\mathrm{g^{(2)}(0)}$ to the sum of $\eta_e$ and $\eta_f$. From fitting the $\ket{2_a,0_b}$ fringes we can extract values for $\eta_d$ and $\eta_e + \eta_f$. For the fit presented in \autoref{fig:NPhotonFringes} of the main manuscript, we fix $\eta_e = 0.32$ so that $\eta_e \sim \eta_f$. We note that these values do not correspond to our measured detector efficiency of $0.82 \pm 0.05$ as they also include the losses associated with fiber coupling at the interferometer output and from the long runs of fiber with multiple connections between the experiment and the detectors.

For the $\ket{1_a,1_b}$ fit, we fix the mode efficiencies using the values obtained from the $\ket{2_a,0_b}$ fit. This assumption holds as these measurements were taken consecutively, without any adjustment to the interferometer. As $\mathrm{g^{(2)}(0)}$ and $\mathcal{I}$ influence the fringes in the same way, they cannot both be free parameters in the fit. Instead, we perform two fits, one where $\mathrm{g^{(2)}(0)}$ is fixed at the independently measured value of $0.018$, and one as an additional cross-check where $\mathcal{I}$ is fixed at the value of 0.974 from the fixed $\mathrm{g^{(2)}(0)}$ fit. The fit with fixed $\mathcal{I}$ reproduces the independently measured $\mathrm{g^{(2)}(0)}$ value with a small uncertainty, indicating a well-defined set of fitted parameters despite potentially strong mutual dependency.

Finally, for the $\ket{2_a,2_b}$ fringes we again fix the values of $\eta_c = 0.803$, $\eta_d = 0.761$ and $\eta_f = 0.322$ derived from the previous fits. We also choose to fix the source parameters $\mathrm{g^{(2)}(0)} = 0.018$ and $\mathcal{I} = 0.974$ as these values have been verified in previous fits. The resulting $\eta_e = 0.178 \pm 0.037$ is now significantly lower than $\eta_f$, reflecting the additional losses associated with our quasi-PNR three photon detection scheme. The full results of each fit are shown in \autoref{tab:fitvalues}.

\begin{table}[h!]
    \centering
    \begin{tabular*}{\textwidth}{@{\extracolsep{\fill}}lcccccc}
        \toprule
        Input & $\mathrm{g^{(2)}(0)}$ & $\mathcal{I}$ & $\eta_c$ & $\eta_d$ & $\eta_e$ & $\eta_f$ \\
        \midrule
        $\ket{1_a0_b}$ & $0.000^\dagger$ & N/A* & $0.803^\dagger$ & \makecell[c]{$0.781$ \\ $\pm 0.006$} & 1.00$^\dagger$ & N/A* \\
        $\ket{2_a0_b}$ & $0.018^\dagger$ & N/A* & $0.803^\dagger$ & \makecell[c]{$0.761$ \\ $\pm 0.009$} & \makecell[c]{$0.320^\dagger$} & \makecell[c]{$0.322$ \\ $\pm 0.070$} \\
        $\ket{1_a1_b}_{g^{(2)}(0)^{\dagger}}$ & $0.018^\dagger$ & \makecell[c]{$0.974$ \\ $\pm 0.007$} & $0.803^\dagger$ & $0.761^\dagger$ & $0.320^\dagger$ & $0.322^\dagger$ \\
        $\ket{1_a1_b}_{\mathcal{I}^{\dagger}}$ & \makecell[c]{$0.018$ \\ $\pm0.004$} & $0.974^\dagger$ & $0.803^\dagger$ & $0.761^\dagger$ & $0.320^\dagger$ & $0.322^\dagger$ \\
        $\ket{2_a2_b}$ & $0.018^\dagger$ & $0.974^\dagger$ & $0.803^\dagger$ & \makecell[c]{$0.761^\dagger$} & \makecell[c]{$0.178$ \\ $\pm0.037$} & $0.322^\dagger$ \\
        \bottomrule
    \end{tabular*}
    
    \begin{flushleft}
    \footnotesize
    $^\dagger$Denotes parameters which were fixed for a given fit. \\
    $^*$Denotes parameters not considered/ undefined parameters in the fit of the fringes.
    \end{flushleft}
    \caption{Fitting parameters for each mixed input state.}
    \label{tab:fitvalues}
\end{table}

\subsection{Other Possible Four-Photon Coincidences}
\begin{figure}
    \centering
    \includegraphics[width=1\linewidth]{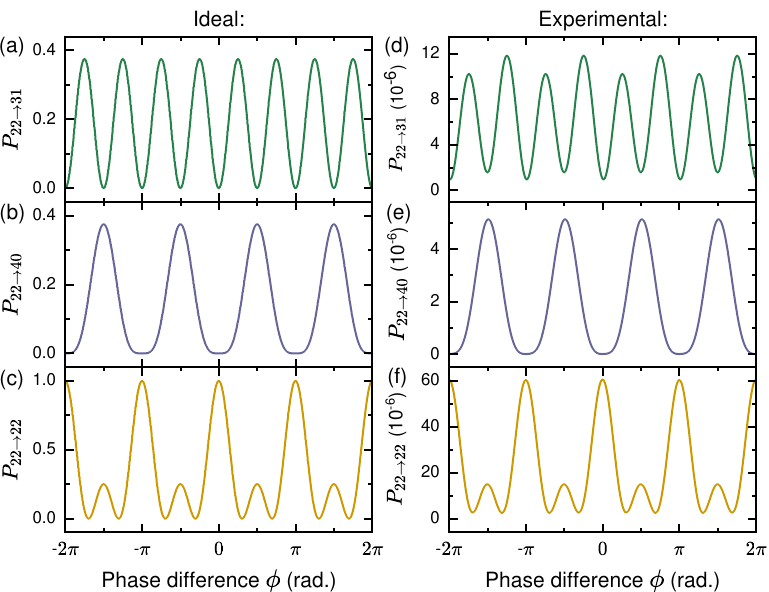}
    \caption{Calculated four-photon interference fringes from the input output model for ideal (a-c) and experimental (d-f) parameters when detecting (a,d) $\ket{3_e1_f}$, (b,e) $\ket{4_e0_f}$ and (c,f) $\ket{2_e2_f}$ output states. $\ket{1_e3_f}$/$\ket{0_e4_f}$ outputs are identical to $\ket{3_e1_f}$/$\ket{4_e0_f}$ aside from rescaling of amplitudes according to the efficiencies $\eta_e/\eta_f$ (for experimental parameters), and are therefore omitted for clarity. \label{Fig:4PhotonPerms}}
\end{figure}

In the main text we focus our investigations of four-photon interference on the $\ket{3_e1_f}$ detection scheme. For completeness, \autoref{Fig:4PhotonPerms} plots the fringes calculated by our model for the other possible permutations of a four-photon coincidence measurement ($\ket{4_e0_f}$,$\ket{2_e2_f}$) for both ideal (\autoref{Fig:4PhotonPerms})(a-c)) and the experimental parameters found by fitting the data in \autoref{fig:NPhotonFringes}(d) of the main manuscript (\autoref{Fig:4PhotonPerms})(d-f)).  From these plots, we observe that $\ket{3_e1_f}$ is the only detection scheme that exhibits consistent four-photon super-resolution across the full range of $\phi$, with the other detection permutations exhibiting non-sinusoidal shapes even for ideal parameters. It is interesting to observe that for integer $\pi$ values of $\phi$, there is a strong peak in $P_{22 \rightarrow 22}$, reaching unity for ideal parameters (\autoref{Fig:4PhotonPerms}(c)). These correspond to the minima that are unaffected by reduced photon indistinguishability ($\mathcal{I}$) in \autoref{fig:NPhotonFringes}(f) of the main text. The origin of this phenomenon is discussed in detail in the main text.

\subsection{Fisher Information Calculations}
To quantify the relationship between the physical parameters of our photon source and the phase sensitivity of our measurements, we employ the method outlined in \cite{Okamoto_2008} which we reproduce here for completeness. In this approach, the phase sensitivity depends both on the raw value and the derivative of the "phase estimate" as a function of phase difference. The phase estimate is defined as the number of times an $N$-photon event is observed in $k$ trials:
\begin{equation}
    C_k(\phi) = kp(\phi)
    \label{equ:ck}
\end{equation}
Where $p(\phi)$ is the phase dependent detection probability for a given output state, which is obtained using our input-output model. The variance of \autoref{equ:ck} is:
\begin{equation}
    \Delta C_k^2 = kp(1-p)
    \label{equ:ckvar}
\end{equation}
The phase error $\delta\phi$ is then defined as:
\begin{equation}
    \delta\phi^2 = \frac{\Delta C_k^2}{\left(\frac{dC_k}{d\phi}_{\phi = 0}\right)^2}
    \label{equ:phaseerror}
\end{equation}
Finally, the phase sensitivity $\mathcal{S}$ is given by the ratio of the phase error to $\frac{1}{\sqrt{kN}}$:
\begin{equation}
    \mathcal{S}^2 = \frac{1}{kN\delta\phi^2}
    \label{equ:phasesen}
\end{equation}
If $\mathcal{S} > 1$ then the standard quantum limit is broken with the Heisenberg limit being reached at a value of $\mathcal{S} = \sqrt{N}$ and the maximum measurable phase sensitivity being $\mathcal{S} = \sqrt{\eta_iN}$. $\mathcal{S}^2$ is also equivalent to the Fisher information per photon.
\begin{figure}[t]
    \centering
   \includegraphics[width=1\linewidth]{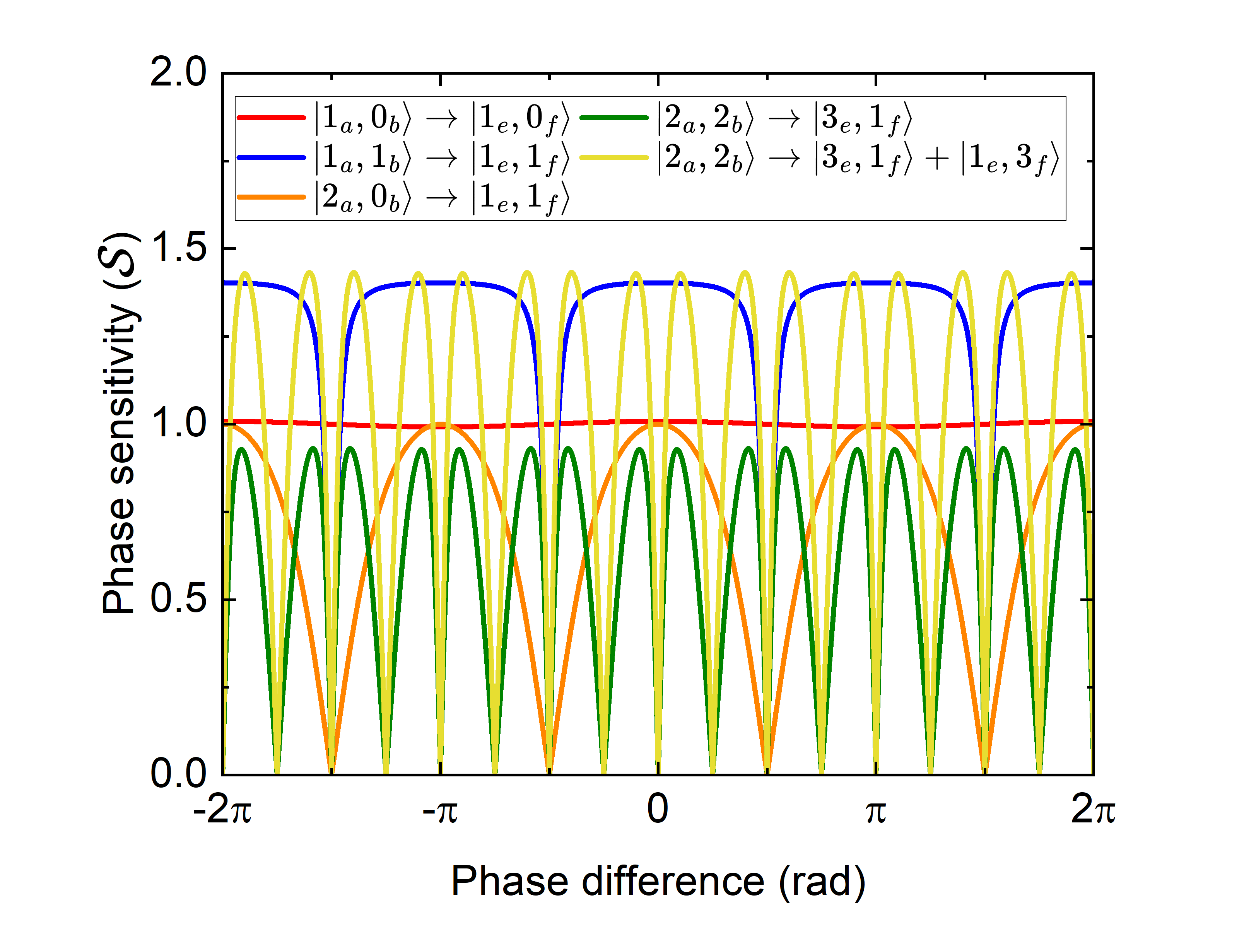}
    \caption{Typical phase sensitivity fringes extracted from our input-output model and \autoref{equ:phasesen}. The parameters used to obtain these fringes are: $\mathrm{g^{(2)}(0)} = 0.016$, $\mathcal{I} = 1$, $\eta_c = \eta_d = \eta_e = \eta_f = 1$ }
    \label{fig:phasefringes}
\end{figure}
To calculate the maximum Fisher information per photon as a function of $\mathrm{g^{(2)}(0)}$ and $\mathcal{I}$, as shown in \autoref{fig:paramsweeps} (b) and (d), we calculate the interference fringes using our input-output model and apply these to \autoref{equ:phasesen}. For each parameter set, the maximum value of the phase sensitivity is extracted from the resulting fringes, an example of which are shown in \autoref{fig:phasefringes}. For the $\ket{1_a0_b}$ and $\ket{1_a1_b}$ inputs, at values of $\phi \sim \pi$, the calculated value of $\mathcal{S}\rightarrow\infty$. These unphysical artefacts are removed before the maximum phase sensitivity is determined for a given fringe.

\subsection{Deriving the relationship between $\mathcal{I}$ and temporal separation}

In \autoref{fig:G2andTdep}(b) of the main text, the relationship between the temporal separation of two photons arriving at the beamsplitter in modes $a$ and $b$ and the measured interference contrast ($C_{11}$) is shown. It is clear that $C_{11}$ has a non-linear relationship with $\mathcal{I}$, as: $C_{11} \rightarrow 1$ when $\mathcal{I} \rightarrow 1$ while $C_{11} \rightarrow0.33$ when $\mathcal{I}\rightarrow0$. This makes extracting the value of $\mathcal{I}$ from the experimental measurements non-trivial. To do so, we must consider the meaning of temporal distinguishability in this context.

We assume $\mathcal{I}$ is equal to the overlap between the temporal wavefunctions of the two photons. We quantify this by calculating the overlap integral between the two wavefunctions:
\begin{equation}
   \mathcal{I} =  \int^{\infty}_{-\infty}\psi^*_1(t) \psi_2(t-\tau)dt
   \label{equ:overlapintegral}
\end{equation}
Where $\psi_1(t)$ and $\psi_2(t)$ are the temporal wavefucntions of the two photons, respectively, and $\tau$ is the separation between their time of arrival on the beamsplitter. 

We can assume that $\psi_1(t)=\psi_2(t)$, as they are produced consecutively by the same single-photon source. From this we obtain the expected behavior that when $\tau = 0$, $\mathcal{I} = 1$ and the photons are temporally indistinguishable, and when $\tau \rightarrow\infty$, $\mathcal{I}\rightarrow0$.

As our QD source is driven by pulsed excitation, the shape of the temporal wavefunction $\psi_1(t)$ will be affected by both the lifetime of the QD decay $\left( T_1\right)$, and laser excitation pulse width $\left(w_p\right)$. We therefore choose to model the temporal wavefunctions as a convolution of the exponential QD decay signal, and the Gaussian laser pulse. 
\begin{figure}
    \centering
    \includegraphics[width=1\linewidth]{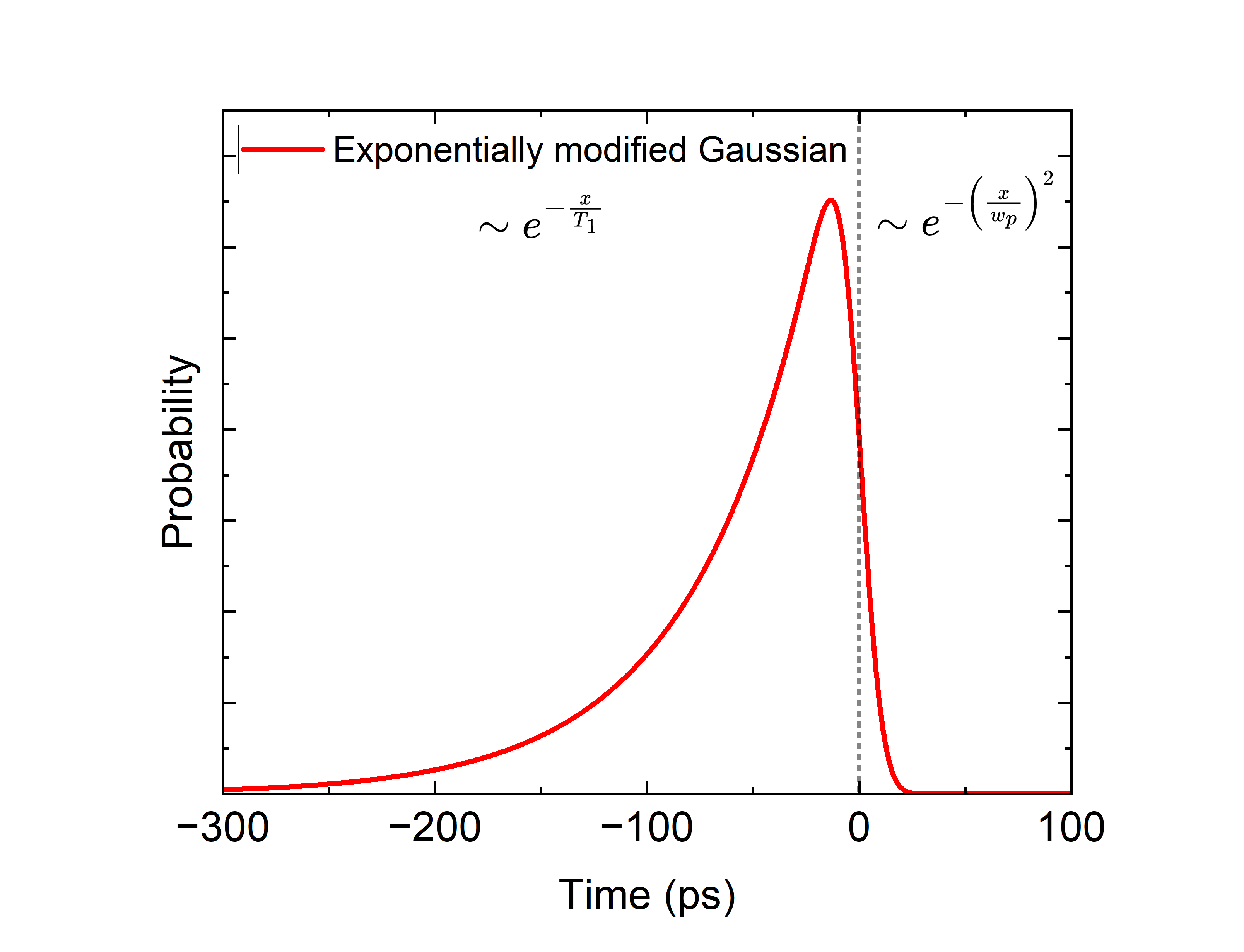}
    \caption{Modelled probability distribution of the single-photon wavefunction. The form of the equation is a convolution between an exponential decay and a Gaussian distribution.}
    \label{fig:ModGaussWVP}
\end{figure}
The resulting function is a exponentially modified Gaussian distribution:
\begin{equation}
    f(x,K) = \frac{1}{2K}\text{exp}\left({\frac{1}{2K^2}-\frac{x}{K}}\right)\text{erfc}\left(\frac{x - \frac{1}{K}}{\sqrt{2}} \right)
    \label{equ:ModGausseq}
\end{equation}
Where $K = \frac{T_1}{w_p}$ and $\text{erfc}(x)$ is the complimentary error function:
\begin{equation}
    \text{erfc}(x) = 1-\text{erf}(x) = \frac{2}{\sqrt{\pi}}\int^\infty_xe^{-t^2}dt
    \label{equ:erfc}
\end{equation}
The resulting shape of the temporal wavefunction $\psi_1(t)$ is shown in \autoref{fig:ModGaussWVP}. The leading edge of the wavefunction is defined by the width of the excitation pulse, as the photon can only be emitted after the pulse has arrived at the QD. The falling edge is defined by the lifetime of the exciton state in the QD. 

By using \autoref{equ:overlapintegral} with $\psi_1(t) = \psi_2(t)$ defined as in \autoref{equ:ModGausseq}, a conversion from temporal separation ($\tau$) to $\mathcal{I}$ is possible. To verify this approach, we construct a fitting function with $T_1 = 59$ ps as a fixed parameter, obtained from an independent measurement of the QD lifetime, and $w_p$ as a free parameter. By using the LMfit module in Python, we obtain a value of $w_p = 8.86\pm 0.29~\mathrm{ps}$. This fitted value is comparable to the nominal $\sim 5~\mathrm{ps}$ pulse duration of our excitation laser, noting that the pulses are likely stretched in time due to filtering by the cavity modes in our cavity-mediated excitation scheme \cite{Javadi_2023}. The resulting conversion between interference contrast $C_{11}$ and $\mathcal{I}$ is presented in the inset to \autoref{fig:G2andTdep}(b) in the main text.

\section{Calibration of Interferometer Phase Shift}
\label{sec:phasecalib}
\begin{figure}[ht]
    \centering
    \includegraphics{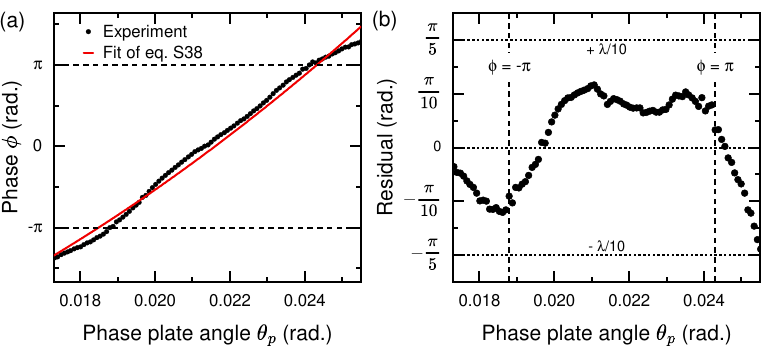}
    \caption{(a) Solid points - experimental phase $\phi$ extracted from \autoref{eq:PhaseIntensity} plotted against the phase plate angle $\theta_p$, solid red line - fit of \autoref{eq:PhasePlateAngle} to the experimental data. Horizontal dashed lines indicate the bounds of the $\pm \pi$ phase range used in the experiments of \autoref{fig:NPhotonFringes} in the main manuscript. (b) Residuals of the fit of \autoref{eq:PhasePlateAngle} presented in (a). Dotted horizontal lines indicate the $\lambda/10$ surface flatness specification of our phase plate, whilst vertical dashed lines again indicate the $\pm \pi$ phase range used in the experiments of \autoref{fig:NPhotonFringes} in the main manuscript.}
    \label{fig:PhaseCalib}
\end{figure}

When tilting the phase plate in the interferometer by an angle $\theta_p$ with respect to normal incidence, we theoretically expect a phase shift $\phi$ according to:

\begin{equation}
    \phi = \frac{n x_t}{\lambda}(1-\mathrm{sec(\theta_p)}), 
\end{equation}
where $x_t$ and $n$ are respectively the thickness and refractive index of the phase plate, $c$ is the speed of light and $\lambda$ is the photon wavelength. Taking the small angle approximation, this becomes:

\begin{equation}
    \phi = \frac{n x_t\theta_p^2}{2\lambda},
    \label{eq:PhasePlateAngle}
\end{equation}
suggesting that we would expect a quadratic relationship between the phase plate angle and interferometer phase.

However, as discussed in the main text, the finite surface flatness ($\lambda / 10$) of the phase plate used in the interferometer implies an unknown relationship between $\theta_p$ and $x_c$ due to spatial fluctuations in thickness. This can lead to slight anharmonicity of the fringes when plotted directly against the phase plate angle, making accurate comparison with theory more challenging. To address this, we use the single-photon fringe data to calibrate the relationship between $\theta_p$ and $\phi$. We normalise the intensity ($I$) of our single-photon fringes into the range $[-1,1]$. Then, applying the fundamental interferometric principle of a sinusoidal relationship between phase and intensity (as captured in \autoref{eq:P10} of the main text), we evaluate the phase according to:

\begin{equation}
    \phi = \mathrm{acos(I)},
    \label{eq:PhaseIntensity}
\end{equation}
with appropriate phase shifts of $\pi$ to avoid duplicated values. This mapping between $\theta_p$ and $\phi$ is used to produce the values of $\phi$ plotted on the horizontal axis of \autoref{fig:NPhotonFringes}.

\autoref{fig:PhaseCalib}(a) shows the experimental phase extracted from \autoref{eq:PhaseIntensity} plotted as solid black points, alongside a fit of \autoref{eq:PhasePlateAngle} (solid red line). The dashed lines at $- \pi$/$\pi$ indicate the boundaries of the region corresponding to the horizontal axis of \autoref{fig:NPhotonFringes} in the main text. It is noticeable that the experimental points deviate from the theoretical curve, however these deviations are small and appear as smooth fluctuations.

To gain further insight, \autoref{fig:PhaseCalib}(b) plots the residuals from the fit shown in \autoref{fig:PhaseCalib}(a). We observe that these residuals fall within the bounds of the $\lambda/10$ surface flatness (dotted horizontal lines) specified for our phase plate, supporting our interpretation that the fluctuations in phase as a function of plate angle (\autoref{fig:PhaseCalib}(a)) originate from imperfect surface flatness of the phase plate. Within the region considered by our experiments in \autoref{fig:NPhotonFringes} of the main text (vertical dashed lines), the residuals fall well within this bound, indicating that the maximum shift arising from our calibration procedure is around $\pi/10$.